\documentclass[12pt,iop,revtex4]{emulateapj}
\usepackage{graphicx,adjustbox}
\usepackage{amssymb}
\usepackage{epstopdf}
\usepackage{textcomp,txfonts} 
\usepackage{dpfloat}

\PassOptionsToPackage{hyphens}{url}
\usepackage[breaklinks=True]{hyperref}

\shortauthors{Pineda et al.}
\shorttitle{Brown Dwarf H$\alpha$ Emission}

\begin{document}
\title{A Survey for H$\alpha$ Emission from Late L dwarfs  and T dwarfs \footnotemark[$\dagger$]}
\author{J. Sebastian Pineda\altaffilmark{1}, Gregg Hallinan\altaffilmark{1}, J. Davy Kirkpatrick\altaffilmark{2}, Garret Cotter\altaffilmark{3}, Melodie M. Kao\altaffilmark{1}, and Kunal Mooley\altaffilmark{3}}

\affil{\altaffilmark{1} California Institute of Technology, Department of Astronomy, 1200 E. California Ave, Pasadena CA, 91125, USA; \href{mailto:jspineda@astro.caltech.edu}{jspineda@astro.caltech.edu}  }
\affil{\altaffilmark{2} Infrared Processing and Analysis Center, MS 100-22, California Institute of Technology, Pasadena, CA 91125, USA}
\affil{\altaffilmark{3} University of Oxford, Department of Physics, Denys Wilkinson Building, Keble Road, Oxford, OX1 3RH, United Kingdom}

\footnotetext[$\dagger$]{The data presented herein were obtained at the W.M. Keck Observatory, which is operated as a scientific partnership among the California Institute of Technology, the University of California and the National Aeronautics and Space Administration. The Observatory was made possible by the generous financial support of the W.M. Keck Foundation.}

\begin{abstract}

Recently, studies of brown dwarfs have demonstrated that they possess strong magnetic fields and have the potential to produce radio and optical auroral emissions powered by magnetospheric currents. This emission provides the only window on magnetic fields in the coolest brown dwarfs and identifying additional benchmark objects is key to constraining dynamo theory in this regime. To this end, we conducted a new red optical (6300 - 9700 \AA) survey with the Keck telescopes looking for H$\alpha$ emission from a sample of late L dwarfs and T dwarfs. Our survey gathered optical spectra for 29 targets, 18 of which did not have previous optical spectra in the literature, greatly expanding the number of moderate resolution (R$\sim$2000) spectra available at these spectral types. Combining our sample with previous surveys, we confirm an H$\alpha$ detection rate of $9.2\pm^{3.5}_{2.1}$~\% for L and T dwarfs in the optical spectral range of L4 - T8. This detection rate is consistent with the recently measured detection rate for auroral radio emission from \cite{Kao2016}, suggesting that geometrical selection effects due to the beaming of the radio emission are small or absent. We also provide the first detection of H$\alpha$ emission from 2MASS 0036+1821, previously notable as the only electron cyclotron maser radio source without a confirmed detection of H$\alpha$ emission. Finally, we also establish optical standards for spectral types T3 and T4, filling in the previous gap between T2 and T5.

\end{abstract}


\section{Introduction}

\begin{deluxetable*}{l c c  c c c c c}
\tablecaption{Observing Log 
\label{tab:obslog} }
\tablehead{ \colhead{Object} & \colhead{SpT NIR/Opt\tablenotemark{a}} & \colhead{UT Date} & \colhead{Filter} & \colhead{Instrument} & \colhead{$t_{\mathrm{exp}}$ (s)} & \colhead{Air Mass} & \colhead{References\tablenotemark{b}}}
\startdata
SDSS J000013.54+255418.6  & T4.5/ T5 & 2014 Dec 22& OG550 &DEIMOS & 2400 &1.18 - 1.25&  15, 5 \\
2MASS J00361617+1821104\tablenotemark{c} & L3.5/L3.5 & 2012 Jul 19 & Clear & LRIS &  5400 & 1.00 - 1.10 & 16/13, 17 \\
SIMP J013656.5+093347.3   & T2.5/ T2 & 2014 Dec 22 & OG550 &DEIMOS & 3600 &1.01-1.10 & 1 \\
			  & 	&    2014 Aug 27& OG550 &LRIS  & 2400 & 1.02 - 1.03 & \\
2MASS J02431371-2453298	  & T6/ T5.5 & 2014 Dec 22 & OG550 &DEIMOS & 2400 & 1.47-1.52 & 4, 5 \\
SDSS J042348.57-041403.5  & T0/ L7.5 & 2014 Dec 22 & OG550 &DEIMOS & 1800 & 1.10  & 11, 5 \\
2MASS J05591914-1404488   & T4.5/ T5 & 2014 Dec 22 & OG550 &DEIMOS & 2400 & 1.21-1.23 & 3, 5  \\
WISEP J065609.60+420531.0& T3/ T2     & 2014 Dec 22 & OG550 &DEIMOS & 2400 & 1.08 & 14 \\ 
2MASS J07003664+3157266\tablenotemark{d} & -- / L3.5 & 2014 Feb 03& GG400 &DEIMOS & 1200 & 1.24  & 20 \\
2MASS J07271824+1710012    & T7/ T8  & 2014 Feb 03 &GG400 & DEIMOS & 1800 & 1.27 & 4, 5 \\
SDSS J075840.33+324723.4  & T2/ T3 & 2014 May 05 &GG495 & DEIMOS & 2400 & 1.21 - 1.28  & 15, 5  \\
WISE J081958.05-033528.5 & T4/ T4 & 2014 May 05 & GG495 &DEIMOS & 2400 & 1.09-1.10 & 14 \\
2MASSI J0835425-081923	  & -- / L5 & 2014 Feb 03& GG400 &DEIMOS & 1200 & 1.35 & 7 \\
SDSSp J092615.38+584720.9 & T4.5/ T5 & 2014 Dec 22 & OG550 &DEIMOS & 2400 & 1.35-1.38 & 11, 5 \\
2MASS J09393548-2448279   & T8/ T8 & 2014 Feb 03 & GG400 &DEIMOS & 2000 & 1.57 & 19, 5 \\ 
2MASS J10430758+2225236	  & -- / L8 & 2014 Dec 22 & OG550 &DEIMOS & 2400 & 1.03-1.06 & 8 \\
SDSS J105213.51+442255.7  & T0.5/ L7.5 & 2014 May 05 & GG495 &DEIMOS & 3600 & 1.10-1.12 & 6 \\
2MASS J11145133-2618235	  & T7.5/ T8 & 2014 Feb 03 & GG400 &DEIMOS & 2000 & 1.47 & 19, 5 \\
2MASS J12314753+0847331	  & T5.5/ T6 & 2014 Dec 22 &OG550 & DEIMOS & 2400 & 1.18-1.25 & 15, 5 \\
SDSS J141624.08+134826.7  & L6/ L6 & 2014 May 05 & GG495 & DEIMOS & 1800 & 1.05 & 2 \\
WISEP J150649.97+702736.0& T6/ T6 &  2014 May 05 & GG495 & DEIMOS & 2000 & 1.42 & 14 \\
2MASSW J1507476-162738    & L5.5/ L5 & 2014 May 05& GG495 &DEIMOS & 900 & 1.25 & 16/13, 15\\
SDSSp J162414.37+002915.6 & T6/ T6 & 2014 May 05 &GG495 & DEIMOS & 1800 & 1.07 & 18, 5 \\
PSO J247.3273+03.5932\tablenotemark{e}     & T2/ T3 & 2014 May 05 & GG495 &DEIMOS & 1800 & 1.12 & 9 \\
WISEP J164715.59+563208.2 & L9p/ L7 & 2014 May 05 & GG495 &DEIMOS & 2000 & 1.42 & 14 \\
2MASS J17502484-0016151   & L5.5/ L5 & 2014 May 05 & GG495 &DEIMOS & 900 & 1.07 & 12 \\
2MASS J17503293+1759042	  & T3.5/ T4 & 2014 May 05 &GG495 & DEIMOS & 1200 & 1.07 & 11, 5  \\
2MASS J17545447+1649196  & T5.5/ T5.5 & 2014 May 05 & GG495 &DEIMOS & 1800 & 1.11 & 16 \\
2MASS J21392676+0220226	  & T1.5/ T2 & 2014 Aug 27 & OG550 &LRIS & 2400 & 1.07 - 1.09 & 10 \\ 
2MASS J22541892+3123498   & T4/ T5  & 2014 Dec 22  & OG550 &DEIMOS & 2400 & 1.12-1.17 & 4, 5
\enddata
\tablenotetext{a}{Optical spectral types are from this paper. For NIR spectral types see references.}
\tablenotetext{b}{The first entry is the discovery reference and second the NIR spectral type reference unless otherwise noted.}
\tablenotetext{c}{2MASS J00361617+1821104 - optical spectral type from \cite{Kirkpatrick2000}; observations were taken with the LRIS Dichroic, D560.}
\tablenotetext{d}{2MASS J07003664+3157266 - optical spectral type from \cite{Thorstensen2003}.}
\tablenotetext{e}{RA, = 16 29 18.409, DEC = +03 35 37.10 .}
\tablenotetext{}{R\textsc{eferences.} -- (1) \citealt{Artigau2006}; (2) \citealt{Bowler2010}; (3) \citealt{Burgasser2000b}; (4) \citealt{Burgasser2002};  (5) \citealt{Burgasser2006}; (6) \citealt{Chiu2006};  (7) \citealt{Cruz2003}; (8) C07;   (9) \citealt{Deacon2011}; (10) \citealt{Faherty2012};  (11) \citealt{Geballe2002};  (12) \citealt{Kendall2007}; (13) \citealt{Kirkpatrick2000};  (14) \citealt{Kirkpatrick2011}; (15) \citealt{Knapp2004}; (16) \citealt{Reid2000}; (17) \citealt{Reid2001} ; (18) \citealt{Strauss1999}; (19) \citealt{Tinney2005};  (20) \citealt{Thorstensen2003}}
\end{deluxetable*}

Twenty years ago, the discovery of the first brown dwarfs opened up the study of sub-stellar objects as interesting astrophysical targets spanning the gap between stars and planets \citep{Nakajima1995,Oppenheimer1995}. Since then, our understanding of brown dwarfs has developed considerably, including their atmospheric properties, evolution and internal structure (e.g., \citealt{Burrows2001, Kirkpatrick2005,Marley2015} and references therein). Many of these developments are based on detailed spectroscopic analyses examining brown dwarf spectra at infrared (IR) wavelengths, where the photospheric flux is the brightest, and where the effects of absorption bands, such as CH$_{4}$, H$_{2}$O and NH$_{3}$, are most prominent (e.g., \citealt{Burgasser2002, Burgasser2006b, Burgasser2006, Burgasser2010, McLean2003, Knapp2004, Cushing2005,Cushing2008,Cushing2011, Stephens2009, Kirkpatrick2012, Mace2013}). By comparison, the flux at red optical wavelengths (600 nm - 1000 nm) is much fainter because of the cool effective temperatures, especially for late L dwarfs, T dwarfs and Y dwarfs ($T_{\mathrm{eff}}\,<\,1500 $ K). Consequently, there have been far fewer studies looking at cool brown dwarfs at these wavelengths.

One feature of particular interest in the optical spectrum is H$\alpha$ emission at 6563 \AA, which has often been used as an indicator of chromospheric emission in the spectra of M dwarfs and early L dwarfs (e.g., \citealt{Reiners2008, Schmidt2015}). For early M dwarfs, the chromospheric nature of the Balmer emission is substantiated by accompanying evidence in X-ray and UV data, revealing high temperature atmospheric regions consistent with a transition region between the photosphere and a corona (e.g., \citealt{Linsky1982,Fleming1988,Walkowicz2008}). For late M dwarfs, L dwarfs and cooler objects, despite very few detections in the X-ray and UV, the presence of chromospheres has been inferred in the population based on detections of H$\alpha$ emission and the analogy with the warmer stars. However, X-ray and optical observations of ultracool dwarfs (UCD; spectral type $\ge$M7) show a drop in their X-ray and H$\alpha$ luminosities \citep{Berger2010, Williams2014,Schmidt2015}. The decline in these X-ray and H$\alpha$ emissions, has been seen as indicative of a decline in the ability of UCDs to sustain much magnetic activity in their cool atmospheres with a transition taking place around the boundary between stars and brown dwarfs (\citealt{Mohanty2002, Reiners2008, Berger2010}).

Despite the cool atmospheric temperatures, recent surveys have revealed that many brown dwarfs also show strong radio emission, indicating that they can sustain strong magnetic fields throughout the whole spectral sequence from L dwarfs to T dwarfs \citep{Hallinan2008, Berger2010, Route2012, Williams2014, Kao2016}. Early efforts to understand this radio emission invoked standard Solar-like magnetic processes \citep{Berger2006,Berger2009,Route2012}. Building on those efforts, continued monitoring of the radio brown dwarfs and observations of the coolest UCDs have now shown that the pulsed radio emission is entirely consistent with being a consequence of the electron cyclotron maser instability (ECMI) as part of auroral currents in the magnetosphere \citep{Hallinan2008,Hallinan2015,Williams2015,Lynch2015,Kao2016}. Furthermore, \cite{Hallinan2015} demonstrated that certain optical spectral features, Balmer series emission lines and broadband variability, can be directly tied to the auroral process in some objects. The connection is further corroborated by the association of radio aurorae with H$\alpha$ emission, suggested by a high detection rate of radio brown dwarfs in the late L dwarf and T dwarf regime, when selecting the observational sample based on potential auroral activity indicators \citep{Kao2016}. This survey is in stark contrast to the very low detection rate of numerous previous surveys that looked for radio emission from brown dwarfs \citep{Berger2006, McLean2012, Antonova2013, Route2013}.

Within the UCD regime, many objects may indeed exhibit chromospheric emissions, especially for the late M dwarfs and early L dwarfs, in which the atmospheres are warmest. However, many studies have now shown that auroral processes are also possible throughout the brown dwarf sequence. This leads to the questions: what governs brown dwarf magnetic activity and what drives potential auroral activity? Amongst late M dwarfs and early L dwarfs, disentangling the different processes requires dedicated monitoring of known benchmark objects like 2MASS~0746+2000 and LSR~1835+3259 \citep{Berger2009, Hallinan2015}.Another way to examine the question is to observe late L dwarfs and T dwarfs, objects in which the local stochastic heating of the upper atmosphere that generates chromospheric emission, as seen on the Sun, is difficult to generate. A study of the activity in these objects allows us to understand the prevalence of magnetic processes, assess the viability of the auroral mechanism and find new potential benchmark targets for dedicated monitoring. If the H$\alpha$ emission in these objects is associated with the same processes that produce auroral radio emission, surveys in the optical provide an additional means to look for brown dwarfs potentially harboring auroral activity. The advantage of surveying for magnetic activity in the optical, over the radio, is that at radio wavelengths the emission may be highly beamed, as in the auroral case, and thus a detection can be highly dependent on the viewing geometry of the system, but is less dependent on geometry at optical wavelengths \citep{Treumann2006}. These factors motivate the search for H$\alpha$ emission, potentially of an auroral nature, in the optical spectrum. We note that optical variability due to weather phenomena is also a compelling reason to observe and monitor the spectra of objects in this spectral range \citep{Heinze2015}.

Our focus is on late L dwarfs and T dwarfs which have received less followup at optical wavelengths than the warmer brown dwarfs but are much brighter in the optical than the cooler Y dwarfs. Much of these initial efforts took place $\sim$10 years ago, prominently by \cite{Kirkpatrick1999}, \cite{Burgasser2003} and \cite{Cruz2007}, hereafter K99, B03 and C07, respectively. The early studies were not able to get detailed optical spectra for all of the brown dwarfs that emerged from early wide-field infrared sky surveys like the Two Micron All Sky Survey (2MASS; \citealt{Skrutskie2006}). Moreover, since then, numerous all sky surveys, including the Sloan Digital Sky Survey (SDSS; \citealt{York2000}), the DEep Near-Infrared Survey of the Southern Sky (DENIS; \citealt{Epchtein1997} ), the Wide-field Infrared Survey Explorer (WISE; \citealt{Wright2010}), the Panoramic Survey Telescope and Rapid Response System (Pan-STARRS; \citealt{Kaiser2002}), the United Kingdom Infrared Telescope Deep Sky Survey (UKIDDS; \citealt{Lawrence2007}), and the Canada-France Hawaii Telescope Legacy Survey (CFHTLS; \citealt{Delorme2008}), have greatly expanded the number of known late L dwarfs and T dwarfs, many of them bright enough to observe with large ground based telescopes. The growing number of late L dwarfs and T dwarfs, thus allows for a comprehensive assessment of the prevalence of H$\alpha$ activity for objects of this effective temperature range. 

Due to the small number of T-dwarf studies at optical wavelengths, our current observational understanding of their spectra remains defined by the early works, like B03, and the few handful of targets they observed. These early spectra set the optical spectral sequence in this regime and provide the observational archetypes for T dwarf optical spectral features \citep{Kirkpatrick2005}. The prominent features include the pressure-broadened wings of the K \textsc{i} resonant doublet at 7665/7699 \AA, other alkali lines from Cs \textsc{i} and Rb \textsc{i}, as well as molecular band-heads of CrH and H$_{2}$O (B03; \citealt{Kirkpatrick2005}). These features are physically interesting because they are sensitive to temperature, gravity, metallicity and the rainout of cloud condensates \citep{Burrows2002}. An expanded collection of optical spectra will thus allow us to examine these features in greater detail.

In this article, we present the results of a new survey of late L dwarfs and T dwarfs at red optical wavelengths looking for H$\alpha$ emission. In Section~\ref{sec:data}, we discuss our observations and the target selection for our data sample. In Section~\ref{sec:opticalspec}, we present the collection of optical spectra, including literature data and examine the variety of optical spectral features. In Section~\ref{sec:Ha}, we focus in particular on the H$\alpha$ emission and the prevalence of potential auroral activity. In Section~\ref{sec:objects}, we discuss our findings for a series of particularly interesting objects in our data sample. Lastly, in Section~\ref{sec:summary}, we summarize and discuss our results.

\begin{figure*}[htbp]
\begin{leftfullpage}
   \centering
   \includegraphics[height=0.9\textheight]{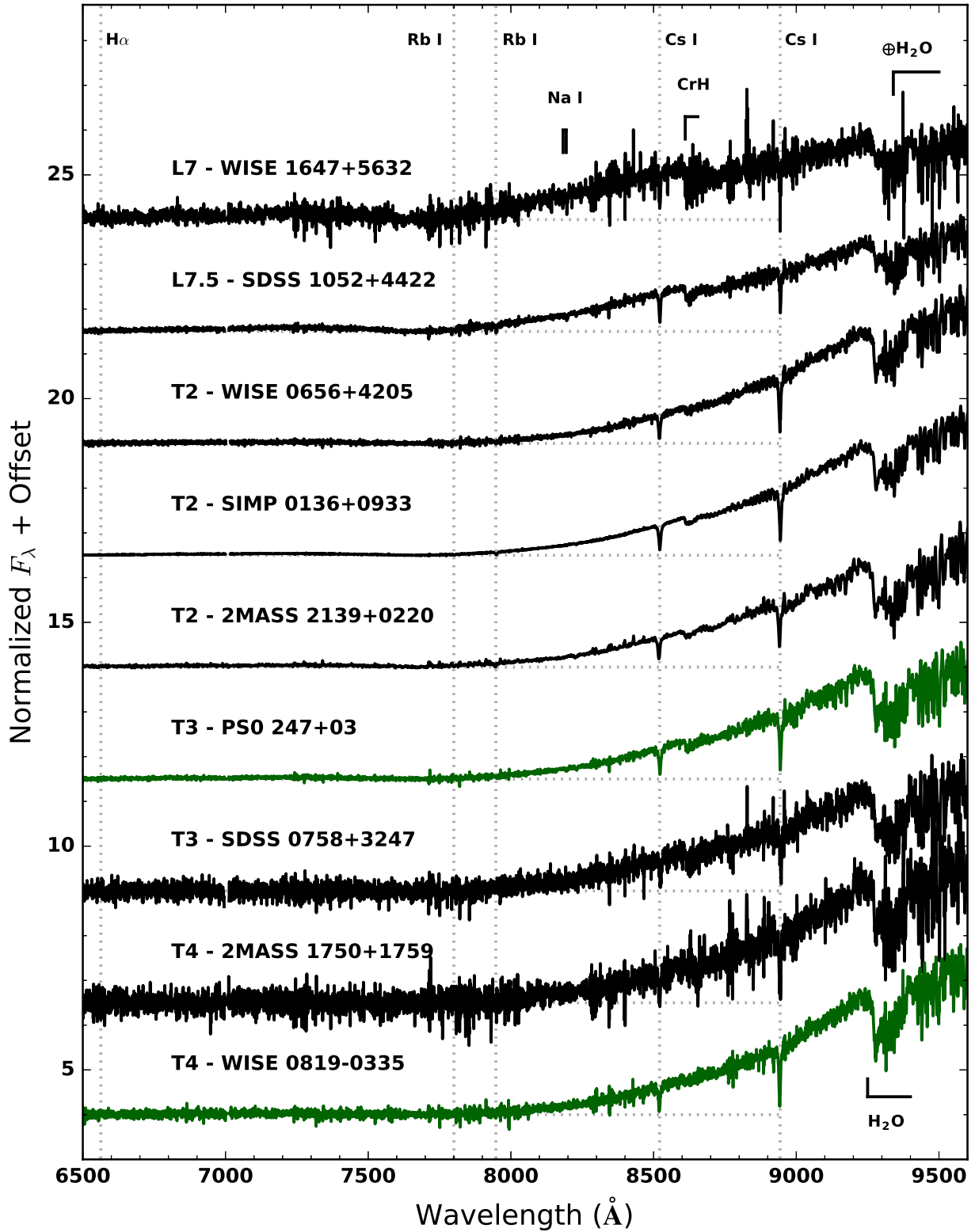} % requires the graphicx package
   \caption{Sequence of spectra for objects in our sample without optical spectra in the literature, arranged according to the optical spectral type. The spectra are normalized at 8750 $\mathrm{\AA}$ with some of the main spectral features noted -- see also Figure~\ref{fig:exSIMP0136}. We also note the spectra of objects PSO 247+03 and WISE 0819-0335 as candidates for the optical T3 and T4 spectral standards (shown in green). Note that there is a slight break in the DEIMOS spectra at 7010 $\mathrm{\AA}$ due to a gap in the DEIMOS detector. All the spectra plotted here were taken with DEIMOS, except for the spectrum of 2MASS J2139+02, which was taken with LRIS. }
   \label{fig:optseq1}
   \end{leftfullpage}

\end{figure*}

\begin{figure*}[htbp]
\begin{fullpage}
   \centering
   \includegraphics[height=0.9\textheight]{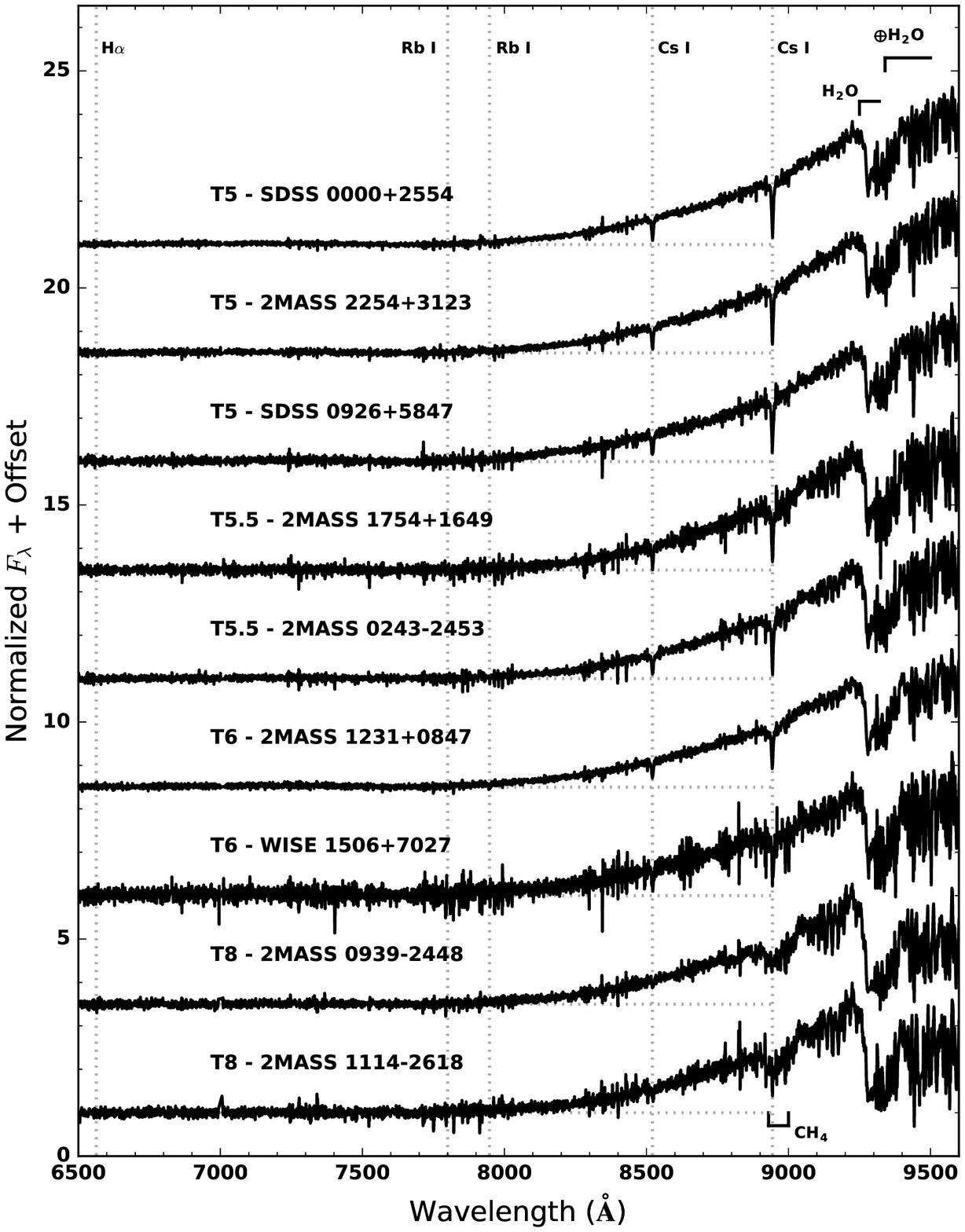} % requires the graphicx package
   \caption{Continued from Figure~\ref{fig:optseq1}}
   \label{fig:optseq2}
   \end{fullpage}
\end{figure*}

\section{Data}\label{sec:data}

\subsection{Sample} 

We selected our target sample by examining the collection of known brown dwarfs in the literature and culling targets that already had observations at red optical wavelengths. We used the compendium of brown dwarfs at \texttt{DwarfArchives.org} as a resource in this endeavor, including new updates to the archive (private communication -- Chris Gelino) and also cross checked the Ultracool RIZzo Spectral Library\footnotemark[1]. We gave priority to the brightest and closest targets, using the Database of Ultracool Parallaxes\footnotemark[2] to verify the distances \citep{Dupuy2012}. We further combined our new observations of these targets with literature T dwarf spectra (discussed in Section~\ref{sec:obs}). All together this resulted in the largest compilation of late L dwarf and T dwarf red optical spectra yet assembled. 

\footnotetext[1]{Data from the Library Available at \url{http://dx.doi.org/10.5281/zenodo.11313}}
\footnotetext[2]{Database accessible at \href{http://www.as.utexas.edu/~tdupuy/plx/Database_of_Ultracool_Parallaxes.html}{http://www.as.utexas.edu/$\sim$tdupuy/plx/Database\textunderscore \\ of\textunderscore Ultracool\textunderscore Parallaxes.html}}

\begin{figure*}[htbp]
   \centering
   \includegraphics[width=0.9\textwidth]{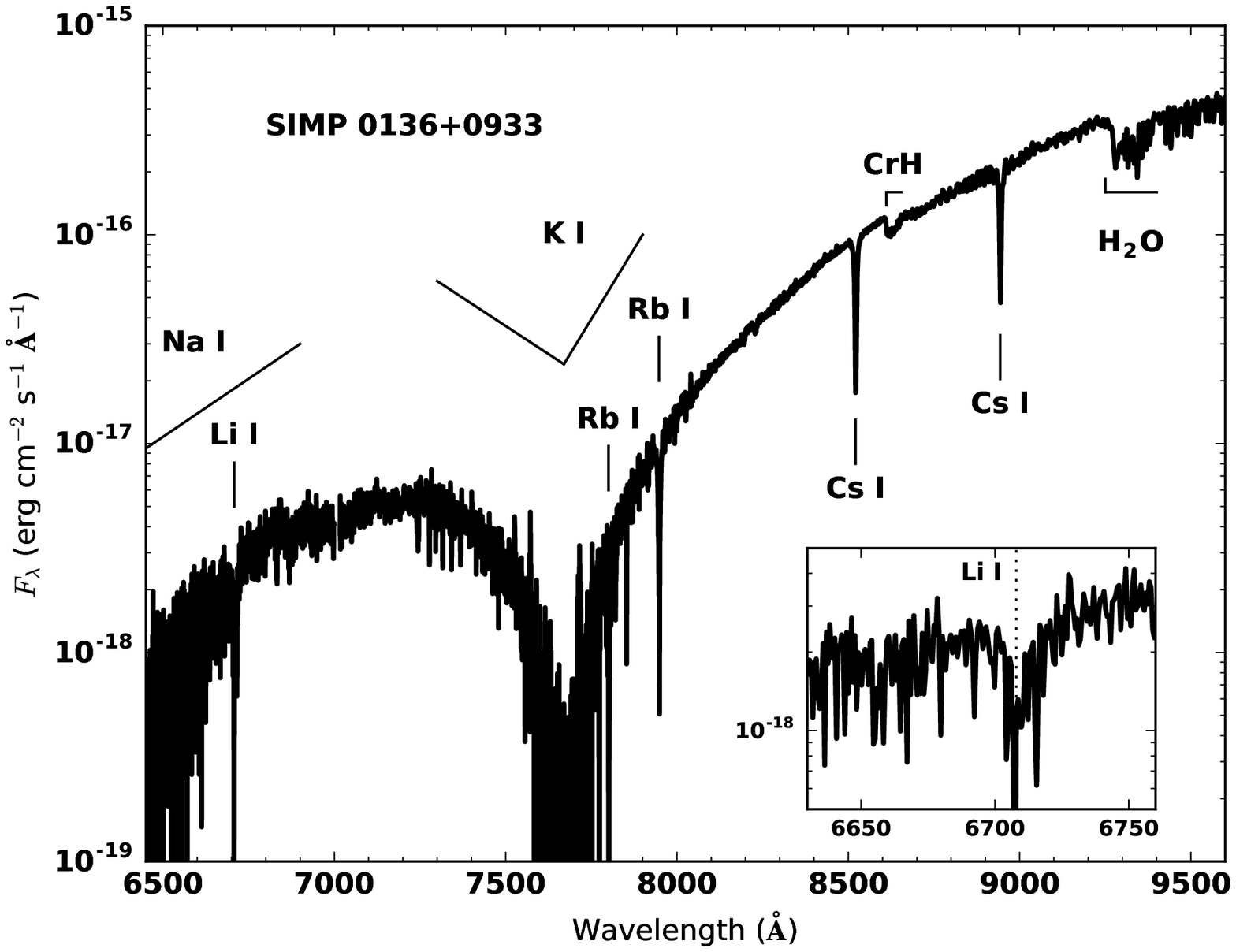} % requires the graphicx package
   \caption{The DEIMOS spectrum of SIMP 0136+0933, a T2 brown dwarf. The figure indicates the location of prominent alkali absorption lines and molecular features that shape the optical spectra of T dwarfs and define the T dwarf optical spectral sequence. The inset shows the spectral region around the Li \textsc{i} line, which is clearly detected in this target -- see Section~\ref{sec:SIMP}. }
   \label{fig:exSIMP0136}
\end{figure*}

\subsection{Observations}\label{sec:obs}

We observed our target brown dwarfs at the W. M. Keck Observatory using either the Low Resolution Imaging Spectrometer (LRIS) on Keck I or the DEep Imaging Multi-Object Spectrograph (DEIMOS) on Keck II during the course of several observing nights, mostly in 2014 \citep{Oke1995, Faber2003}. The observations were designed predominately for the purposes of searching for H$\alpha$ emission in objects without previous observational limits on the emission strength. However, in our survey we also looked at some objects with previous limits, testing for variability, as well as at targets that may have only had marginal detections, which we sought to confirm. The full observing log for the 29 objects is presented in Table~\ref{tab:obslog}. We also display the spectra of the 18 objects without previous optical spectra in Figure~\ref{fig:optseq1} and Figure~\ref{fig:optseq2}.

\subsubsection{DEIMOS}\label{sec:deimos}

The majority of our survey was conducted with Keck/DEIMOS, a multi-slit imaging spectrograph designed for acquiring optical wavelength spectra of faint objects. DEIMOS operates at the Nasmyth focus and includes a flexure compensation system for increased stability. The multi-slit capability utilizes pre-milled masks, however we used the instrument in longslit mode with the standard longslit masks, placing the targets in the 1" (8.4 pixels) wide slit. This mode is regularly used for spectroscopic followup for the Palomar Transient Factory (PTF; \citealt{Law2009}). The detector uses a large format 8k x 8k CCD mosaic, consequently the blue and red ends of a single spectroscopic exposure fall on different CCD chips with a small gap between them. 

For our observations we used the 600 line mm$^{-1}$ grating blazed at 7000  $\mathrm{\AA}$ yielding a wavelength coverage from 5000 - 9700 $\mathrm{\AA}$ and a resolution of 3.5 $\mathrm{\AA}$ (R $\sim$ 2000) with a dispersion of 0.62 $\mathrm{\AA}$ pixel$^{-1}$. The blue extent of the spectra were also limited by the use of order blocking filters to limit the effects of second order light (see Table~\ref{tab:obslog}). In comparison to the earlier observations in K99, B03 and C07, our observations are at a slightly higher resolution.

\footnotetext[3]{The longslit masks that we used have short breaks along the length of the slit to prevent the slitmask from buckling.}

\footnotetext[4]{Python Software Foundation. Python Language Reference, version 2.7.10. Available at \url{http://www.python.org}. PyRAF is a product of the Space Telescope Science Institute, which is operated by AURA for NASA.}

Data reduction for these observations was initiated with a modified version of the DEEP2 pipeline utilized by PTF \citep{Cooper2012,Newman2013}. The pipeline uses the overscan region to bias subtract the raw frames, median combines the dome flats to flat field the raw data, determines the wavelength solution from NeArKrXe arc lamps and returns the 2D spectrum of each slit for both the red and blue CCD chips with cosmic ray rejection routines applied.\footnotemark[3] The rest of the data reduction was handled by custom routines in \texttt{Python} with \texttt{PyRAF} to rectify the frames, extract the spectra and flux calibrate.\footnotemark[4]

Since the ultracool dwarfs have very red spectra, up to a couple orders of magnitude flux difference between 7000 \AA$\;$and 9000 \AA, for most targets the spectral flux is barely seen toward the blue end of the detector ($ \lambda \lesssim7000\; \mathrm{\AA}$). Thus, the extractions for each CCD chip were done independently and we therefore used the location of the centroid of the target trace on the red chip in the rectified frames as the central location for the blue chip. We verified that this produced accurate results based on the calibration targets and the brighter L dwarfs with plenty of flux for $\lambda \lesssim7000\; \mathrm{\AA}$.

Our observations took place on 2014 February 3rd, 2014 May 5th and 2014 December 22nd (UT). At the beginning of the observing night for February  3rd there was fog at the summit, however the dome opened up half way through the night with typical seeing conditions of 1.2". Conditions on May 5th were more favorable, low humidity and 0.8" seeing. December 22nd was also a good observing night with good conditions throughout and 0.7" seeing. Typical exposure times varied between 900 s for the L dwarfs and up to 1800 s for the fainter T dwarfs with multiple exposures for some of the targets. We also observed a standard star from \cite{Hamuy1994} or \cite{Massey1990} each night for the purposes of flux calibrating the spectra. These calibrators were Hiltner 600, HZ44 and Feige 110, respectively for the three nights. We note that the specific order blocking filter varied for each of the three nights, GG400, GG495 and OG550, respectively. 

Although, this did not effect the primary goal of surveying for H$\alpha$ emission, unlike the other filters, the use of the GG400 filter (observations on Feb. 3rd) limited the ability to flux calibrate the red end of the spectrum due to contamination from second order short wavelength light in the standard star observations. Consequently, we had to make adjustments in order to get flux calibrated spectra for that observing night. We took advantage of the fact that a couple of L dwarf targets we observed that evening, were part of the Ultracool RIZzo spectral library. To get a sensitivity curve for the red chip on that night, we divided the raw extracted spectrum by the literature spectra of the same targets, 2MASSI J0835425-081923 and 2MASS J07003664+3157266AB, took the median of the respective curves and fit a low order polynomial. We subsequently scaled the resulting curve to match the sensitivity function from the blue chip where there was no effect from the second order light. The resulting sensitivity curve agreed reasonably well with similar curves from the other observing nights and the reduced spectra from the night of Feb. 3rd proved to match the expected optical standards rather well (see Section~\ref{sec:stands}).

The red end of the DEIMOS spectrum cuts off at $\sim$9700 $\mathrm{\AA}$ in the middle of a broad telluric H$_{2}$O absorption band. This presented an added difficulty in determining the flux calibration for a given night because the telluric band could not be interpolated over when determining the sensitivity function from the spectrophotometric standards. Consequently, the sensitivity function for $\lambda > 9300\; \mathrm{\AA}$ is based solely on the polynomial fit at shorter wavelengths. The effect this has on the spectral shape is only significant beyond $\sim$9400 $\mathrm{\AA}$ where the spectrum is also significantly effected by telluric absorption. Additionally, we did not correct for telluric absorption in any of the DEIMOS spectra. Thus, the effects of the telluric absorption are most prominent in the same region where the flux calibration is most unreliable. This has no impact on the bulk of our analysis as we focus on short wavelength regions, however it does have a small effect on the measurement of the H$_{2}$O feature at 9250 $\mathrm{\AA}$ (see Section~\ref{sec:SPR}; B03).

\subsubsection{LRIS}\label{sec:obslris}

\footnotetext[5]{IRAF, version 2.16, is distributed by the National Optical Astronomy Observatories, which are operated by the Association of Universities for Research in Astronomy, Inc., under cooperative agreement with the National Science Foundation. Available at \url{http://iraf.noao.edu}.}

We observed the spectrum of 2MASS J00361617+1821104 (2MASS~0036+1821) on 2012, July 19th. These observations used the 1200 line mm $^{-1}$ grating blazed at 6400 \AA~ through a 0.7'' slit, yielding a wavelength coverage of 5600-7200 \AA, and a resolution of $\sim1.7$ \AA~($R\sim3700$). The detector was readout with 2x2 binning, yielding a dispersion of 0.81 \AA~pixel$^{-1}$. The target was observed for 5400 seconds split into six 900~s exposures. The observations also used the LRIS dichroic D560 with a clear filter through the red arm of the instrument. We also took data with the blue side of LRIS, however, we do not present that data in this paper. These data were reduced with the \texttt{longslit} routines in IRAF.\footnotemark[5] The individual exposures were bias subtracted; corrected for pixel-to-pixel gain variation and slit illumination via dome flats; transformed onto a rectilinear wavelength-sky position grid via internal arc lamps; and finally sky-subtracted by interpolating a polynomial along each row in the sky direction, excluding the target from the fit by sigma-clipping. 

We acquired LRIS spectra of SIMP J013656.5+093347.3 (SIMP~0136+0933) and 2MASS~J21392676+0220226 (2MASS~2139+0220) on 2014 August 27th. Although LRIS is designed to use a beamsplitter to allow independent and simultaneous observations in a red channel and a blue channel, these observations made use of only the red channel. Since the work of K99 and B03, which used LRIS to obtain optical spectra of late L dwarfs and T dwarfs, the red channel detector has been upgraded, improving the sensitivty \citep{Rockosi2010}.

Our observations on 2014 August 27th used the 400 line mm$^{-1}$ grating blazed at 8500 $\mathrm{\AA}$ through a 0.7" slit, yielding a wavelength coverage of 6300 - 10100 $\mathrm{\AA}$, a resolution of $\sim5$ \AA~($R\sim 1700$), and a dispersion of 1.33 $\mathrm{\AA}$ pixel$^{-1}$. The data were taken through a companion program, and accidentally left out the order blocking filter, which meant the flux calibration was not viable from that night's observing (the dichroic was also set to clear). As we did with the February 3rd DEIMOS observations (see Section~\ref{sec:deimos}), we used the flux calibrated DEIMOS observations of SIMP~0136+0933 from 2014 December 22nd, to calibrate for the rough shape of the LRIS sensitivity function. This did not effect the blue end of the spectrum, nor our ability to measure the H$\alpha$ emission, however it created an effective upper limit to the LRIS wavelength coverage of 9700 $\mathrm{\AA}$. We reduced the data using standard routines in \texttt{PyRAF}. Since the spectral trace becomes very faint in the red, we used the trace for calibration white dwarfs taken before and after the science observations to define the extraction trace for our target brown dwarfs. 

In our sample we also include the archival data of the 7 T dwarfs with red optical spectra from the WISE followup of \cite{Kirkpatrick2011}, in order to bolster the sample size of T dwarf optical spectra and provide comparisons of the optical features across the full optical sequence to the latest T dwarf spectral types. These objects are, the T5, WISE 1841+7000, the T7s, WISE 1019+6529 and WISE 2340-0745, the T8s, WISE 1617+1807, WISE 1457+5815 and WISE 1653+4444, and the T9, WISE 1741+2553. These observations also used Keck/LRIS with similar settings but a wider 1" slit. The reductions and calibrations of those data are described in K99 and \cite{Kirkpatrick2006}. These objects were incorporated into the analysis of the optical spectral features (Section~\ref{sec:features}) but not for H$\alpha$ emission (Section~\ref{sec:Ha}).

We also used the archival spectra of SDSSp J083717.22-000018.3, SDSSp J102109.6-030419, and 2MASS J12095613-1004008 to get additional measurements of T dwarf H$\alpha$ emission \citep{Kirkpatrick2008}. These objects already had measurements of the important optical spectral features (see Section~\ref{sec:features}) but required flux estimates of their H$\alpha$ emission instead of just equivalent widths (see Section~\ref{sec:Ha}).

\begin{turnpage}
\begin{deluxetable*}{r c c  c c c c c c c c c c c} % 14 cols
\tablecaption{Alkali Line Pseudo-Equivalent Widths
\label{tab:alkali} }
\tablehead{ & & & \multicolumn{2}{c}{ 8521 $\mathrm{\AA}$ Cs \small{I}}  && \multicolumn{2}{c}{ 8943 $\mathrm{\AA}$ Cs \small{I} \tablenotemark{a}  }&& \multicolumn{2}{c}{ 7800 $\mathrm{\AA}$ Rb \small{I}} && \multicolumn{2}{c}{ 7948 $\mathrm{\AA}$ Rb \small{I}}\\ \cline{4-5} \cline{7-8}  \cline{10-11} \cline{13-14} \\ 
\colhead{Object} &  \colhead{Opt. Spectral Type} && \colhead{$\lambda_{0}$ (\AA)} & \colhead{pEW (\AA)} && \colhead{$\lambda_{0}$ (\AA)} & \colhead{pEW (\AA)}&& \colhead{$\lambda_{0}$ (\AA)} & \colhead{pEW} (\AA)&& \colhead{$\lambda_{0}$ (\AA)} & \colhead{pEW (\AA)} }
\startdata

2MASS 0700+3157 & L3.5 && 8521.72$\, \pm \,0.02$ & 3.64$\, \pm \,0.04$&& 8943.98$\, \pm \,0.02$ & 2.32$\, \pm \,0.04$&& 7800.96$\, \pm \,0.03$ & 4.62$\, \pm \,0.08$&& 7948.38$\, \pm \,0.03$ & 4.34$\, \pm \,0.06$ \\ [3.5pt]
2MASS 0835-0819 & L5 && 8521.68$\, \pm \,0.01$ & 4.72$\, \pm \,0.03$&& 8943.903$\, \pm \,0.022$ & 2.56$\, \pm \,^{0.04}_{0.03}$&& 7800.79$\, \pm \,0.04$ & 6.51$\, \pm \,0.11$&& 7948.22$\, \pm \,0.02$ & 6.18$\, \pm \,0.06$ \\ [3.5pt]
2MASS 1507-1627 & L5 && 8520.42$\, \pm \,0.01$ & 5.81$\, \pm \, 0.03$&& 8942.596$\, \pm \,0.014$ & 3.41$\, \pm \,0.03$&& 7799.76$\, \pm \,0.04$ & 9.01$\, \pm \,^{0.16 }_{0.15}$&& 7947.01$\, \pm \,0.02$ & 8.13$\, \pm \,0.07$ \\ [3.5pt]
2MASS 1750-0016	& L5 && 8522.00$\, \pm \, 0.01$ & 5.84$\, \pm \,0.04$&& 8944.21$\, \pm \,0.02$ & 3.47$\, \pm \,0.04$ && 7801.20$\, \pm \, 0.05$ & 9.63$\, \pm \,^{0.20}_{0.19}$&& 7948.485$\, \pm \,0.025$ & 8.15$\, \pm \,^{ 0.09}_{ 0.08}$ \\ [3.5pt]
SDSS 1416+1348 & L6 && 8519.99$\, \pm \,0.01$ & 6.43$\, \pm \,0.03$&& 8942.07$\, \pm \,0.01$ & 3.91$\, \pm \,0.03$&& 7799.24$\, \pm \, 0.06$ & 11.11$\, \pm \,0.22 $&& 7946.54$\, \pm \, 0.02$ & 9.13$\, \pm \, 0.08$ \\ [3.5pt]
WISE 1647+5632 & L7 && 8521.6$\, \pm \, 0.2$ & 7.2 $\, \pm \,0.5$&& 8943.8$\, \pm \,0.3$ & 4.2$\, \pm \,0.6$&& . . . & $<42$ && 7947$\, \pm \,2$ & 12$\, \pm \,^{16}_{6}$ \\ [3.5pt]
SDSS 0423-0414 & L7.5 && 8521.81$\, \pm \,0.02$ & 7.84$\, \pm \,0.07$&& 8943.86$\, \pm \, 0.03$ & 5.62$\, \pm \,0.06$&& 7801.0$\, \pm \,0.2$ & 9.3$\, \pm \,^{ 0.7 }_{ 0.6}$&& 7948.30$\, \pm \,0.07 $ & 9.22$\, \pm \,0.20$ \\ [3.5pt]
SDSS 1052+4422 & L7.5 && 8521.96 $\, \pm \,0.06 $ & 6.82$\, \pm \,^{0.17}_{0.16}$&& 8944.27$\, \pm \, 0.08$ & 4.69$\, \pm \, 0.16$&& . . . & $<23 $ && 7948.7$\, \pm \, 0.3 $ & 8.6$\, \pm \,^{ 1.0 }_{ 0.9 }$ \\ [3.5pt]
2MASS 1043+2225 &L8  & & 8520.68$\, \pm \,0.06$ & 7.41$\, \pm \,0.18$&& 8942.77$\, \pm \, 0.07$ & 5.42$\, \pm \,^{ 0.16 }_{ 0.15}$&& . . . & . . .  && 7947.2$\, \pm \,0.3$ & 9.6$\, \pm \, 0.9$ \\[3.5pt]
2MASS 1632+1904 & L8 && $8520.7 \pm  0.1$ & $7.6 \pm 0.3$ && $8942.2 \pm 0.3$ & $5.5 \pm 0.5$ && . . . & . . . && . . . & . . .  \\ [3.5pt]
SDSS 0837-0000 & T0 && $8520.5 \pm  0.3$ & $9.0 \pm^{0.7}_{6}$ && $8942.5 \pm 0.3$ & $8.1 \pm 0.7$ && . . . & . . . && . . . & . . .  \\ [3.5pt]
SIMP 0136+0933 & T2 && 8521.77$\, \pm \,0.02$ & 9.70$\, \pm \,0.06$&& 8943.969$\, \pm \,0.014$ & 8.58$\, \pm \,0.04 $&& 7801.40$\, \pm \,0.27 $ & 10.6$\, \pm \,^{ 1.2 }_{ 1.1}$&& 7948.58$\, \pm \,0.12$ & 9.96$\, \pm \,0.32$ \\ [3.5pt]
	& && 8519.07$\, \pm \,0.02$ & 10.16$\, \pm \,0.06$&& 8941.35$\, \pm \,0.01$ & 8.45$\, \pm \,0.04$&& 7798.8$\, \pm \,0.6$ & 14$\, \pm \,2$&& 7945.81$\, \pm \,0.13 $ & 10.96$\, \pm \,0.38$ \\ [3.5pt]
WISE 0656+4205 & T2 && 8520.52$\, \pm \,0.07$ & 9.57$\, \pm \,0.23$&& 8942.54$\, \pm \,0.05$ & 8.45$\, \pm \,0.15$&& . . . & $<56$ && 7948$\, \pm \,1$ & 12$\, \pm \,3$ \\ [3.5pt]
2MASS 2139+0220 & T2 && 8518.75$\, \pm \, 0.06$ & 8.55$\, \pm \,0.17$ && 8941.11$\, \pm \,0.05$ & 7.39$\, \pm \, 0.12 $&& . . .& . . . && 7944.2$\, \pm \,0.5$ & 10$\, \pm \,1$ \\ [3.5pt]
SDSS 0758+3247 & T3 && 8524.2$\, \pm \,0.3$ & 8.4$\, \pm \,^{ 0.9 }_{ 0.8}$&& 8946.3$\, \pm \,0.2$ & 7.4$\, \pm \,0.6$&& . . . & . . .&& . . . & $<37$ \\ [3.5pt]
PSO 247+03 & T3 && 8522.18$\, \pm \, 0.07$ & 10.23 $\, \pm \, 0.23 $&& 8944.17$\, \pm \, 0.06 $ & 8.45$\, \pm \, 0.16 $&& . . . & $<41$ && 7951$\, \pm \,^{ 2 }_{ 1 }$ & 10$\, \pm \,^{ 5 }_{ 3 }$ \\[3.5pt]
WISE 0819-0335 & T4 && 8519.97$\, \pm \, 0.12$ & 8.97$\pm 0.42 $&& 8942.00$\pm0.10$ & 8.52$\pm0.26$&& . . . & $<51$ && . . .  & $<33$  \\ [3.5pt]
2MASS 1209-1004 & T4 && 8522.6$\, \pm \, 0.3$ & 12.2$\, \pm \, 0.6 $&& 8945.4$\, \pm \, 0.3$ & 9.7 $\, \pm \, 0.5 $ && . . . & . . .&& . . . & $< 10$ \\ [3.5pt]
2MASS 1750+1759 & T4 && 8521.1$\, \pm \,0.3$ & 11.7$\pm^{1.2}_{ 1.1 }$&& 8942.95$\pm0.27$ & 8.5$\pm0.7$&& . . . & . . .&& . . . & $<89$ \\ [3.5pt]
SDSS J0000+2554  & T5 && 8521.48$\, \pm \, 0.07$ & 8.79$\pm0.23$&& 8943.81$\pm0.06$ & 8.38$\pm0.15$&& . . . & $<48$ && 7950$\, \pm \,^{ 2}_{ 1}$ & 10$\, \pm \,^{ 10 }_{3 }$ \\[3.5pt]
2MASS 0559-1404 & T5 && 8521.565$\, \pm \, 0.029$ & 7.73 $\, \pm \,0.10$ && 8943.66$\, \pm \,0.02$ & 7.48$\, \pm \,0.07$&& 7800.1$\, \pm \,0.5$ & 12$\, \pm \,^{3}_{2}$&& 7948.15$\, \pm \,0.25$ & 11.5$\, \pm \,0.7$ \\ [3.5pt]
SDSS 0926+5847 & T5 && 8521.07$\, \pm \,0.15$ & 9.2$\, \pm \,0.4$&& 8943.6$\, \pm \,0.1$ & 8.57$\, \pm \,0.25$&& . . . & $<69$ && . . . & $<32$ \\ [3.5pt]
WISE 1841+7000 & T5 && 8519.$\, \pm \,2.$ & 7$\, \pm \,^{ 4}_{3}$&& 8942$\, \pm \,1.$ & 13.822077$\, \pm \,^{ 4}_{ 3}$ && . . . & . . . && . . . & . . . \\ [3.5pt]
2MASS 2254+3123 & T5 && 8521.29$\, \pm \, 0.10 $ & 8.7$\pm0.3$&& 8943.59$\pm0.07$ & 8.26$\pm0.18$&& . . . & $<108$ && . . . & $<25$ \\ [3.5pt]
2MASS 0243-2453 & T5.5 && 8521.5$\, \pm \,0.1$ & 8.6$\, \pm \,0.4$&& 8943.71$\, \pm \,0.10$ & 8.88$\, \pm \,0.23$&& . . . & . . .&& . . . & $<41$ \\ [ 3.5pt]
2MASS 1754+1649 & T5.5 && 8521.44$\, \pm \,0.17$ & 8.1$\, \pm \,^{0.6}_{0.5}$&& 8943.68$\, \pm \,0.15$ & 8.64$\, \pm \,0.37$&& . . . & . . . && . . . & $<69$ \\ [3.5pt]
2MASS 1231+0847 & T6 && 8521.01$\, \pm \,0.13$ & 7.1$\, \pm \,0.3$&& 8942.93$\, \pm \,0.09$ & 7.46$\, \pm \, 0.21$&& . . . & $<35$ && 7946$\, \pm \,1 $ & 11$\, \pm \,3$ \\[3.5pt]
WISE 1506+7027 & T6 && 8522.19$\, \pm \, 0.28$ & 7.6$\, \pm \, 0.8$&& 8945.0$\, \pm \, 0.3$ & 7.9$\, \pm \,0.6$ && . . . & . . .&& . . . & $<40$ \\ [3.5pt]
SDSS 1624+0029 & T6 && 8520.68$\, \pm \, 0.10$ & 6.65$\, \pm \,^{0.28}_{0.27}$&& 8942.78$\, \pm \,0.09$ & 7.18$\, \pm \,0.21$&& . . . & $<44$ && 7945$\, \pm \,^{ 1 }_{ 2 }$ & 10$\, \pm \,^{ 4}_{3}$ \\ [3.5pt]
WISE 1019+6529 & T7 && . . . & $<19$ && 8944$\, \pm \,^{ 1}_{20}$ & 5$\, \pm \,^{5}_{ 3}$ && . . . & . . . && . . . & . . . \\ [3.5pt]
WISE 2340-0745 & T7 && 8519.9$\, \pm \,0.7 $ & 6$\, \pm \, 1 $&& 8942.4$\, \pm \,0.5$ & 7.4$\, \pm \, 0.8$ && . . . & . . . && . . . & . . . \\ [3.5pt]
2MASS 0727+1710 & T8 && 8521.2 $\, \pm \,0.2 $ & 5.13$\, \pm \,^{0.35}_{ 0.34}$&& 8943.43$\, \pm \,0.15$ & 6.7$\, \pm \,0.3$&& . . . & $<23$ && 7949.1$\, \pm \,^{ 0.5 }_{ 0.6}$ & 8$\, \pm \,2 $ \\ [3.5pt]
2MASS 0939-2448 & T8 && 8521.7$\, \pm \,0.6$ & 2.8$\, \pm \,0.5$&& . . . & . . .&& . . . & . . .&& . . . & $<25$  \\ [3.5pt]
2MASS 1114-2618 & T8 && 8523.1$\, \pm \, 0.5$ & 4.7$\, \pm \,^{ 1.0 }_{0.8}$&& . . . & . . . && . . . & . . .&& . . . & . . . 

\enddata
\tablenotetext{a}{For the L dwarfs the line is confused with a broader molecular absorption band. For the latest T dwarfs the line is obscured by CH$_{4}$ absorption.}
\end{deluxetable*}
\end{turnpage}

\begin{figure}[htbp]
   \centering
   \includegraphics[width=0.5\textwidth]{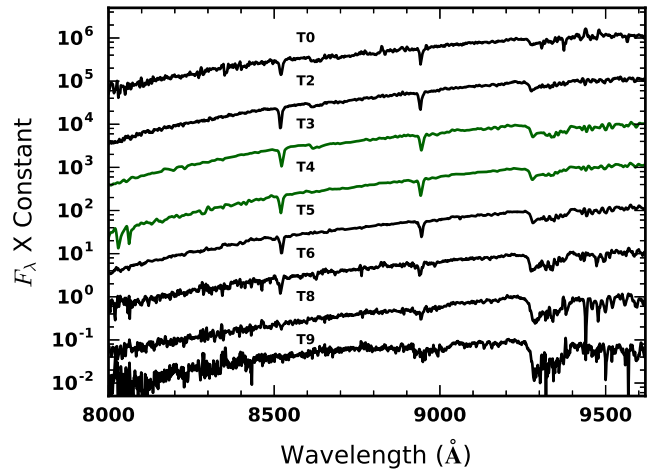} % requires the graphicx package
   \caption{The T dwarf optical spectral standards including our new additions for types T3 and T4. The T0 standard is from \cite{Kirkpatrick2008} and the T9 is from \cite{Kirkpatrick2011}, while the rest are from B03. The T3 and T4 spectra have not been corrected for telluric absorption but have been convolved here to match the resolution of the literature standards.}
   \label{fig:opt_stands}
\end{figure}

\begin{figure}[htbp]
   \centering
   \includegraphics[width=0.45\textwidth]{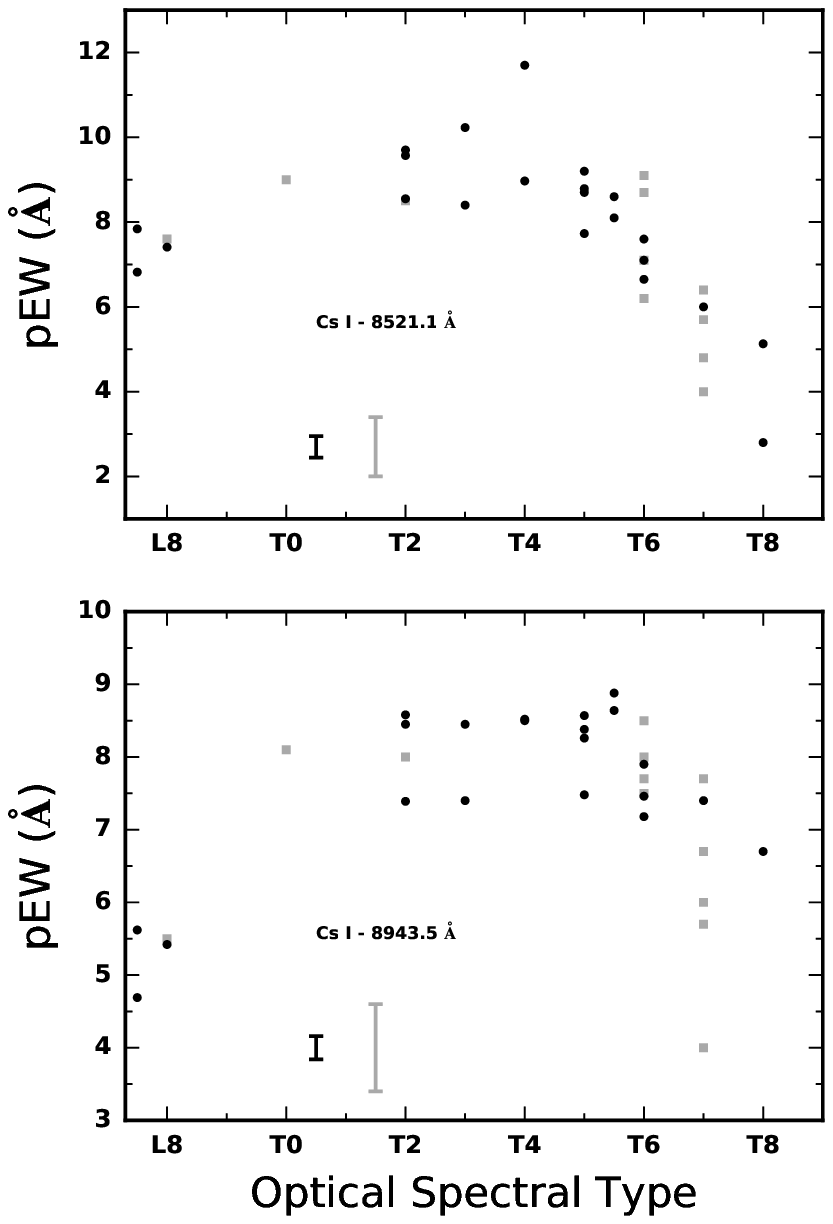} % requires the graphicx package
   \caption{The measured pEWs of the Cs \textsc{i} absorption lines at 8521 $\mathrm{\AA}$ (top) and 8943 $\mathrm{\AA}$ (bottom) as a function of optical spectral type across the T dwarf sequence. We plotted the values listed in Table~\ref{tab:alkali} (black circles), and also the literature measurements from B03 (grey squares), including new measurements for the L8 and T0 standards, 2MASS 1632+1904 and SDSS 0837-0000, respectively. We only plot the values with less than 20\% uncertainties and include the median full errorbar length in the lower left of each plot. The absorption peaks for T2-T4 objects and declines in late T dwarfs. The scatter at a given spectral type is likely associated with differences in gravity and/or metallicity between the different brown dwarfs. The larger scatter for late T dwarfs in the Cs absorption line at 8943 $\mathrm{\AA}$ has to do with the onset of a CH$_{4}$ band around the location of the Cs line.}
   \label{fig:csa_spt}
\end{figure}

\section{Optical Spectra}\label{sec:opticalspec}

\subsection{Spectral Sequence}\label{sec:stands}

To determine the optical spectral types for each object with new optical spectra, we compared the spectra visually with the set of optical spectral standards from L5 to T8. The optical standards are DENIS 1228-1547, for L5 (K99), 2MASS 0850+1057, for L6 (K99), DENIS 0205-1159, for L7 (K99), 2MASS 1632+1904, for L8 (K99),  SDSS 0837-0000, for T0 \citep{Kirkpatrick2008}, SDSS 1254-0122, for T2 (B03), 2MASS 0559-1404, for T5 (B03), SDSS 1624+0029, for T6 (B03), 2MASS 0415-0935, for T8 (B03) and WISE 1741+2553, for T9 \citep{Kirkpatrick2011}. To aid in the visual classification we also convolved the DEIMOS spectra down to the same resolution as the optical standards using a Gaussian kernel. We verified the results of the convolution process by matching our convolved DEIMOS spectra to the literature spectra of the same targets matching the resolution of the standards (ex. 2MASS 0559-1404). The new optical spectral types are included in the Table~\ref{tab:obslog} alongside the near-infrared spectral types from \cite{Burgasser2006} or \cite{Kirkpatrick2011}. We also show the spectra in Figure~\ref{fig:optseq1} and Figure~\ref{fig:optseq2}, with the important spectral features detailed closely in Figure~\ref{fig:exSIMP0136}.

Through this comparison we discovered that four of the spectra display features that are clearly between those of the T2 and T5 spectral standards. The morphology is best illustrated by the strength of the CrH absorption, Cs \textsc{i} lines and the H$_{2}$O absorption. This last water feature is at wavelengths that are influenced by telluric absorption in our spectra, however the astrophysical signal completely dominates (see B03). The objects PSO 247+03 and SDSS 0758+3247 showed slightly weaker CrH absorption relative to the T2 standard and slightly stronger H$_{2}$O absorption but not as strong as the T5 standard while maintaining strong Cs \textsc{i} absorption. Both of these brown dwarfs have NIR spectral types of T2.  2MASS 1750+1759 and WISEP 0819-0335 show no CrH absorption like the T5 standard but with slightly weaker H$_{2}$O and stronger Cs \textsc{i} lines. These two targets have NIR spectral types of T3.5 and T4 respectively. These targets fill the gap in spectral morphologies between T2 and T5 and we propose that PSO 247+03 and WISEP 0819-0335 be considered the optical spectral standards for T3 and T4, respectively. We plot the standards for the T dwarf optical spectral sequence in Figure~\ref{fig:opt_stands}. Using these standards, we also update the optical spectral types of SDSSp~J102109.6-030419 and 2MASS~12095613-1004008 of T3.5 from \cite{Kirkpatrick2008} to T4.

\begin{figure}[htbp]
   \centering
   \includegraphics[width=0.5\textwidth]{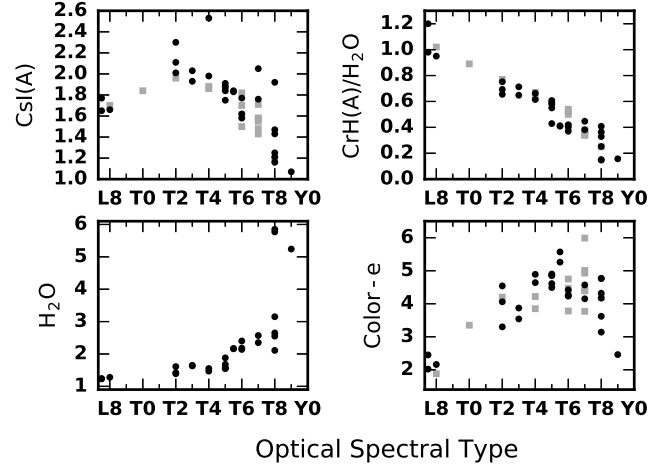} % requires the graphicx package
   \caption{The spectral ratios of Table~\ref{tab:SPR} as a function of optical spectral type. The black points are our new measurements and the grey squares comprise literature values from B03 and \cite{Kirkpatrick2008}. We focus on the T dwarf sequence but show the points down to L7.5 to illustrate how the ratios change across the L/T transition. The CrH(A)/ H$_2$O ratio shows the clearest and tightest trend with spectral type.}
   \label{fig:SPR}
\end{figure}

\subsection{Spectral Features}\label{sec:features}

\subsubsection{Alkali Lines}\label{sec:alkali}

We measured the set of alkali absorption lines in our new spectra. These are the Cs \textsc{i} lines at 8521 $\mathrm{\AA}$  and 8943 $\mathrm{\AA}$  and the Rb \textsc{i} lines at 7800 $\mathrm{\AA}$  and 7948 $\mathrm{\AA}$ (K99). Following B03, to measure the absorption strength of each line we simultaneously fit a Lorentz distribution to the line profile and a linear continuum to the 120 $\mathrm{\AA}$  region centered at the nominal wavelength position of each line. We used a maximum likelihood estimate and a MCMC routine implemented in \texttt{Python} to determine the best fit parameters of the model \citep{Patil2010}. The likelihood was constructed from the product of the probability of each datum which was assumed to be described by a normal distribution centered at the measured flux for each wavelength with the standard deviation given by the error spectrum. The model is given as 

\begin{equation}
S_{\lambda} = b + m(\lambda - \lambda_{0}) + \frac{A}{\pi} \frac{\gamma}{ (\lambda - \lambda_{0})^{2} + \gamma^{2}}  \; ,
\end{equation}

\noindent where $\lambda_{0}$ is the center of the absorption line, $A$ is the total flux absorbed by the line, $\gamma$ defines the width of the Lorentz distribution, $m$ is the slope of the continuum, and $b$ is the level of the continuum at the center of the line. The pseudo-equivalent widths, pEWs, for each line are computed as

\begin{equation}
\mathrm{pEW} = A / b \; .
\end{equation}

Additionally, we compared the corresponding fit using a Voigt line profile (the convolution of a Lorentz profile with a Gaussian profile) to the Lorentz profile fits. Although the Gaussian component is typically dominant in the core whereas the Lorentz profile dominates in the wings of the line from pressure broadening, we found that in all cases the fit using the Voigt line profile model tended to the Lorentz profile with little to no contribution from the Gaussian component. Gaussian line profile fits also did a poor job of fitting the data compared to the Lorentz profile. Additionally, there is a systematic bias in the measured pEWs depending on the assumed shape of the line profile. An assumed Gaussian profile yields lower pEWs than the corresponding fit using a Lorentz line profile with differences of up to 15\%. These results signify the importance of pressure broadening in determining the shape of the absorption line profiles for these high gravity atmospheres.

We report the pEWs for our line profile fits in Table~\ref{tab:alkali}, where we include, in addition to 28 of the targets we observed from Table~\ref{tab:obslog} (all except 2MASS 0036+1821), measurements for 3 of the 7 WISE T dwarfs (see Section~\ref{sec:obslris}) for which we could get decent line fits, and measurements for the L8, T0 and T6 optical standards (using the literature spectra). The T2 and T8 optical standards already have alkali line fitting measurements from B03. Not every line was visible in every spectrum due to the lower flux in the fainter parts of the spectrum. If a line was clearly identified, we fit the line profile as described above. If we detected the continuum but were not able to distinguish a clear absorption line, we determined a 3$\sigma$ upper limit from the uncertainty in the continuum and the sum of the residuals in a 40 $\mathrm{\AA}$ region around the line center after the linear continuum was subtracted. When there was no clear continuum we left the entry in Table~\ref{tab:alkali} blank. All included, the table includes 34 distinct targets. In Figure~\ref{fig:csa_spt}, we plot the pEWs of the Cs \textsc{i} lines as a function of optical spectral type for the T dwarfs with black circles representing our new measurements and the grey squares representing literature data. The peak Cs \textsc{i} absorption across the T dwarf sequence occurs for mid T dwarf spectral types.

\begin{deluxetable}{r c c c c c } % 5 cols
\tablecaption{Spectral Ratios\tablenotemark{a}
\label{tab:SPR} }
\tablehead{ \colhead{Object} & \colhead{SpT Opt} & \colhead{CsA} & \colhead{ H$_{2}$O} & \colhead{CrH/H$_{2}$O} & \colhead{Color-e}}
\startdata

2MASS 0700+3157  &  L3.5  &  1.3  &  1.08  &  1.37  &  1.71 \\ 
2MASS 0835$-$0819  &  L5  &  1.41  &  1.12  &  1.72  &  1.56 \\ 
2MASS 1507$-$1627  &  L5  &  1.56  &  1.12  &  1.79  &  1.56 \\ 
2MASS 1750$-$0016  &  L5  &  1.53  &  1.18  &  1.5  &  1.56 \\ 
SDSS 1416+1348  &  L6  &  1.61  &  1.14  &  1.58  &  1.78 \\ 
WISE 1647+5632  &  L7  &  1.63  &  1.29  &  1.1  &  1.72 \\ 
SDSS 1052+4422  &  L7.5  &  1.65  &  1.23  &  0.98  &  2.45 \\ 
SDSS 0423$-$0414  &  L7.5  &  1.77  &  1.24  &  1.2  &  2.02 \\ 
2MASS 1043+2225  &  L8  &  1.66  &  1.28  &  0.95  &  2.16 \\ 
SIMP 0136+0933  &  T2  &  2.3  &  1.39  &  0.752  &  4.06 \\ 
WISE 0656+4205  &  T2  &  2.01  &  1.42  &  0.693  &  4.54 \\ 
2MASS 2139+0220  &  T2  &  2.11  &  1.61  &  0.64  &  3.8 \\ 
SDSS 0758+3247  &  T3  &  1.93  &  1.63  &  0.713  &  3.54 \\ 
PSO 247+03  &  T3  &  2.03  &  1.65  &  0.648  &  3.87 \\ 
WISE 0819$-$0335 &  T4  &  1.98  &  1.55  &  0.615  &  4.89 \\ 
2MASS 1750+1759  &  T4  &  2.53  &  1.47  &  0.663  &  4.64 \\ 
SDSS J0000+2554  &  T5  &  1.91  &  1.68  &  0.55  &  4.85 \\ 
2MASS 0559$-$1404  &  T5  &  1.75  &  1.59  &  0.579  &  4.61 \\ 
SDSS 0926+5847  &  T5  &  1.88  &  1.55  &  0.59  &  4.49 \\ 
WISE 1841+7000  &  T5  &  1.87  &  1.88  &  0.429  &  4.87 \\ 
2MASS 2254+3123  &  T5  &  1.84  &  1.55  &  0.607  &  4.9 \\ 
2MASS 0243$-$2453  &  T5.5  &  1.83  &  2.17  &  0.409  &  5.26 \\ 
2MASS 1754+1649  &  T5.5  &  1.84  &  2.16  &  0.412  &  5.57 \\ 
2MASS 1231+0847  &  T6  &  1.58  &  2.14  &  0.42  &  4.42 \\ 
WISE 1506+7027  &  T6  &  1.77  &  2.4  &  0.37  &  4.23 \\ 
SDSS 1624+0029  &  T6  &  1.62  &  2.18  &  0.404  &  4.26 \\ 
WISE 1019+6529  &  T7  &  2.05  &  2.57  &  0.447  &  4.57 \\ 
WISE 2340$-$0745  &  T7  &  1.76  &  2.35  &  0.381  &  4.15 \\ 
2MASS 0727+1710  &  T8  &  1.43  &  2.65  &  0.329  &  4.32 \\ 
2MASS 0939$-$2448  &  T8  &  1.21  &  5.85  &  0.147  &  4.77 \\ 
2MASS 1114$-$2618  &  T8  &  1.25  &  5.77  &  0.151  &  4.76 \\ 
WISE 1457+5815  &  T8  &  1.47  &  2.55  &  0.364  &  3.62 \\ 
WISE 1617+1807  &  T8  &  1.92  &  3.15  &  0.252  &  3.14 \\ 
WISE 1653+4444  &  T8  &  1.16  &  2.11  &  0.408  &  4.17 \\ 
WISE 1741+2553  &  T9  &  1.07  &  5.24  &  0.157  &  2.46 
\enddata
\tablenotetext{a}{The spectral ratios are ratios of the flux in the spectrum describing the strength of different spectral features, see Table 5 of B03 for definitions.}
\end{deluxetable}

\subsubsection{Spectral Ratios}\label{sec:SPR}

We also examined the series of spectral ratios summarized by B03 in their Table~5, in particular Cs\textsc{i}(A), CrH(A), H$_{2}$0, and Color-e. The ratios measure the respective spectral features indicated, with Color-e corresponding to the overall spectral slope of the pseudo-continuum between 8450 \AA~ and 9200 \AA. Before measuring the features, we convolved down our DEIMOS spectra to the same resolution as their LRIS sample, as in Section~\ref{sec:stands}, in order to compare their measurements with ours. We also used the line center measurements from the alkali line fitting (Section~\ref{sec:alkali}) to shift each spectrum to a consistent frame in line with the expected positions of the absorption features. The results are presented in Table~\ref{tab:SPR}, where we show measurements from 28 of our newly observed targets, all except 2MASS 0036+1821 for which the spectrum did not cover the selected spectral regions, plus the ratios measured for the 7 late T dwarfs in the the literature from WISE (see Section~\ref{sec:obslris}). We also plot the ratios as a function of optical spectral type in Figure~\ref{fig:SPR}, focusing on the T dwarf sequence with our newly expanded sample.

As demonstrated by B03, the ratio of the CrH(A) feature to the H$_{2}$O feature tracks the T dwarf optical spectral sequence most clearly (see Figure~\ref{fig:SPR}, top right). With our expanded data sample, we show that the relation is rather tight throughout the whole T dwarf optical sequence, despite the influence of weak telluric absorption in the H$_{2}$O index in our new data (See Section~\ref{sec:stands}). A quadratic fit to the spectral types as a function of CrH(A)/H$_{2}$O yields a residual scatter that is less than 1 full spectral type. This particular ratio is the best predictor of the overall spectral morphology, whereas features like the overall optical slope or the alkali line depths show considerably more scatter. Moreover, this ratio combination continues smoothly across the L/T transition.

The H$_{2}$O feature, shown in the lower left of Figure~\ref{fig:SPR} grows gradually through the optical spectral sequence before greatly increasing for spectral types after T8. Two of our targets with new optical spectra, 2MASS 1114$-$2618 and 2MASS 0939$-$2448, showed absorption in line with the T9 optical standard WISE 1741+2553 from \cite{Kirkpatrick2011}. These spectra match the overall shape of the T8 standard, however the H$_{2}$O band is slightly stronger and agrees well with the T9 standard. Because the overall shape so closely matches the T8 standard, we retain the T8 optical spectral type for these objects, however they likely represent a transition to cooler objects, T9s and even Y dwarfs. These remarks are in line with the results of \cite{Burgasser2006b}, which determine that these two objects have effective temperatures cooler than the T8 standard with an upper limit of $T_{\mathrm{eff}} \lesssim 700$ K.

\begin{figure}[htbp]
   \centering
   \includegraphics{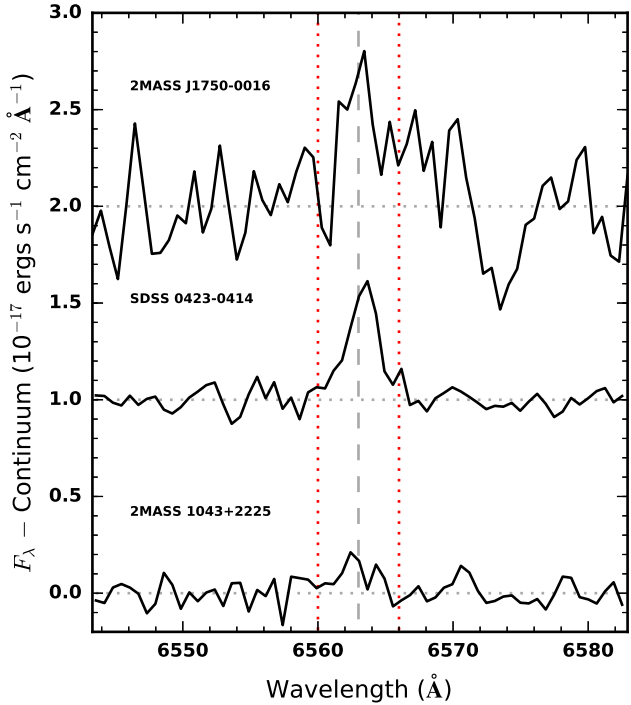} % requires the graphicx package
   \caption{Zoom in of the region around the 6563 $\mathrm{\AA}$ H$\alpha$ emission line (vertical dashed line) for the spectra of 2MASS J1750-0016, SDSS 0423-0414 and 2MASS 1043+2225. The spectra have the local continuum subtracted and are offset by a constant for clarity with the line center marked by a dashed line. The vertical dotted lines delineate the region used to sum the H$\alpha$ flux. We report the measurements of these fluxes in Table~\ref{tab:Halpha}.}
   \label{fig:Ha_prof}
\end{figure}

\begin{figure}[htbp]
   \centering
   \includegraphics{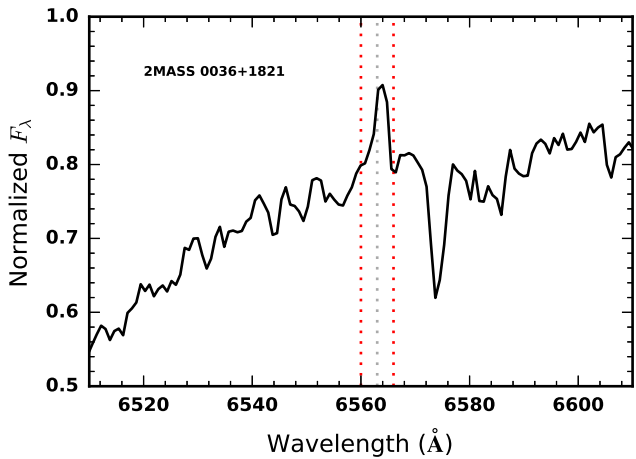} % requires the graphicx package
   \caption{The normalized spectrum of 2MASS~0036+1821 in the region around H$\alpha$. The dashed line marks the expected position of the emission line and the dotted lines denote the region used to add up the emission line flux once the continuum has been subtracted. Unlike the observations for the targets in Figure~\ref{fig:Ha_prof}, the spectrum for 2MASS~0036+1821 was not flux calibrated.}
   \label{fig:M0036}
\end{figure}

\section{H$\alpha$ Activity}\label{sec:Ha}

Of particular interest in this study is the prevalence of H$\alpha$ emission in late L dwarfs and T dwarfs. Most of the spectra did not show a clear indication of H$\alpha$ emission, see Table~\ref{tab:Halpha}. From our observations, only 2MASS~0036+1821, 2MASS~1750$-$0016, SDSS~0423$-$0414 and 2MASS~1043+2225 had excess emission around the location of H$\alpha$. We plot the corresponding H$\alpha$ profiles in Figure~\ref{fig:Ha_prof} and in Figure~\ref{fig:M0036}. To measure this flux we fit a line to the 40~$\mathrm{\AA}$ region around the nominal location of the emission line, excluding the 6 $\mathrm{\AA}$ region centered at 6563 $\mathrm{\AA}$. We subtracted the linear fit from the spectrum and summed the flux between 6560 and 6566 $\mathrm{\AA}$ as the flux in the emission line. The uncertainty was determined from the error spectrum for the sum of that region with the uncertainty in the continuum below the line added in quadrature. In Table~\ref{tab:Halpha}, we report the flux measurements with the 1$\sigma$ uncertainty level as well as the 3$\sigma$ upper limits.

For these measurements we did not apply the line fitting procedure developed in Section~\ref{sec:alkali} for a couple of reasons. Firstly, This approach allowed us to compare our measurements to the values in the literature in a consistent way. Secondly, the line fitting assumes a particular shape for the line profile, which is justified for the absorption lines but not for these emission lines. We did attempt to fit example profiles, Gauss, Lorentz and Voigt, however one did not particularly outperform the others. 

\footnotetext[6]{The new bolometric corrections are consistent with previous efforts by \cite{Liu2010}}

We calculated the ratio of the luminosity in H$\alpha$ to the brown dwarf's bolometric luminosity, $L_{\mathrm{H\alpha}}/L_{\mathrm{bol}}$, by making use of new bolometric corrections from \cite{Filippazzo2015}, which uses the newly defined absolute magnitude scale ($M_{\odot} = +4.74$)\footnotemark[6]. We use the J band bolometric correction as a function of spectral type to determine the bolometric luminosity since it has the least amount of scatter for the field T dwarfs \citep{Filippazzo2015}. The values of $L_{\mathrm{H\alpha}}/L_{\mathrm{bol}}$ for our observations are also listed in Table~\ref{tab:Halpha}. We have also compiled the literature measurements from \cite{Burgasser2000}, \cite{Burgasser2002b}, and B03, and report them in Table~\ref{tab:HalphaL}, with updated values of $L_{\mathrm{H\alpha}}/L_{\mathrm{bol}}$ based on the new bolometric corrections. For three of the objects with literature measurements shown in Table~\ref{tab:HalphaL}, we took new spectra in our current survey with H$\alpha$ measurements shown in Table~\ref{tab:Halpha}: 2MASS 0559$-$1404, SDSS 1624+0029, and 2MASS 0727+1710. Table~\ref{tab:HalphaL} also includes flux measurements for SDSSp J083717.22-000018.3, SDSSp J102109.6-030419, and 2MASS J12095613-1004008, based on archival spectra, which we reanalyzed to provide new limits on the H$\alpha$ flux.

\begin{figure*}[htbp]
   \centering
   \includegraphics[width=0.85\textwidth]{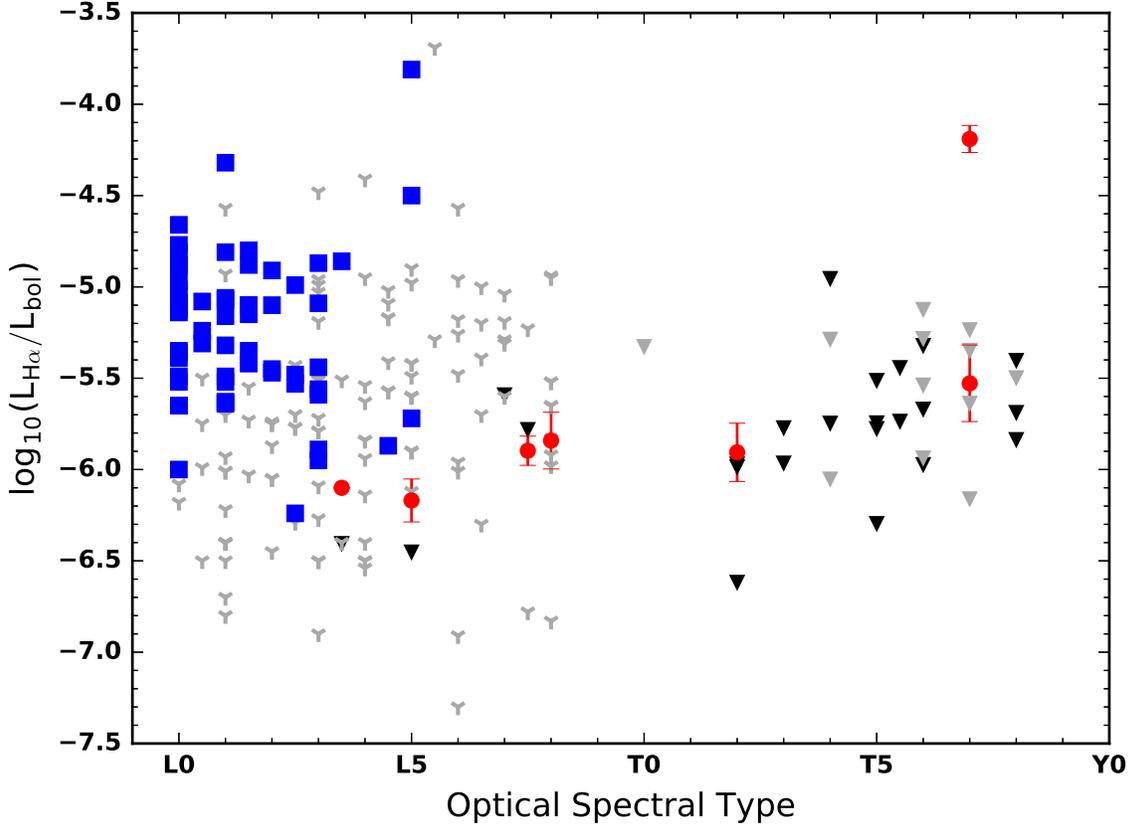} % requires the graphicx package
   \caption{The logarithm of the luminosity in H$\alpha$ relative to an object's bolometric luminosity as a function of optical spectral type. The red points mark measurements listed in Table~\ref{tab:Halpha} and Table~\ref{tab:HalphaL}. Downward triangles mark the upper limits of the same tables with the darker points corresponding to new measurements from this paper and the lighter ones to previous studies of T dwarfs from \cite{Burgasser2000} and B03. Literature values for L dwarf emissions as compiled by \cite{Schmidt2015} and supplemented by new measurements from \cite{Burgasser2015} are included as squares and tri symbols respectively for measurements and upper limits.}
   \label{fig:LHaLbol}
\end{figure*}

\begin{figure}[htbp]
   \centering
   \includegraphics[width=0.45\textwidth]{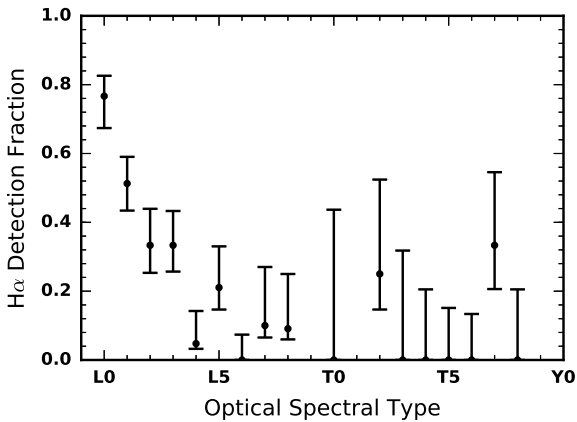} % requires the graphicx package
   \caption{The detection fraction of objects as a function of optical spectral type from L dwarfs to T dwarfs based on the compilations by \cite{Schmidt2015},\cite{Burgasser2015}, and this paper. Half spectral types have been rounded down to earlier spectral types and the errorbars represent the 68\% confidence interval of the corresponding binomial distribution. At spectral types later than about L4/L5 the prevalence of H$\alpha$ detections is low.}
   \label{fig:DetectFrac}
\end{figure}

In Figure~\ref{fig:LHaLbol}, we plot $L_{\mathrm{H\alpha}}/L_{\mathrm{bol}}$ as a function of optical spectral type. Measurements in Table~\ref{tab:Halpha} and Table~\ref{tab:HalphaL} are plotted as red filled circles with new limits as black downward triangles and limits from the literature as grey downward triangles. In the instances in which there were multiple measurements for a single target, either from our observations or in the literature, we plotted a detection, if available, or the most stringent limit for the non-detections. For comparison with earlier spectral types, we have also included measurements (blue squares) and limits (grey tri-symbols) compiled by \cite{Schmidt2015} and supplemented by \cite{Burgasser2015}. These values include measurements from K99, \cite{Kirkpatrick2000}, \cite{Hall2002}, \cite{Liebert2003}, \cite{Schmidt2007}, \cite{Reiners2008}, \cite{Burgasser2011}, and \cite{Schmidt2015}. We do not distinguish binaries in this plot, but note that for those objects, the optical spectrum is dominated by the warmer component and thus the points are representative of the position corresponding to the primary. 

In Figure~\ref{fig:DetectFrac}, we plot the fraction of objects shown in Figure~\ref{fig:LHaLbol} that have H$\alpha$ detections as a function of optical spectral type from L dwarfs through to T dwarfs. Since the data comes from a variety of sources and surveys with different sensitivity limits there are observational biases inherent to this detection fraction. Additionally, many brown dwarfs have been demonstrated to exhibit variability in their H$\alpha$ emission, potentially from rotation \citep{Berger2009,Hallinan2015} or longer timescales (see Section~\ref{sec:0036}). Thus, objects with only non-detections may yet display emission from further monitoring and/or more sensitive observations, so the detection fractions of Figure~\ref{fig:DetectFrac} are systematically low. With these caveats, our extended brown dwarf sample allows us to assess the prevalence of the H$\alpha$ emission, going from L dwarfs to T dwarfs. The data in Figure~\ref{fig:DetectFrac} demonstrates that the declining prevalence of H$\alpha$ emission, demonstrated for early-to-mid L dwarfs in \cite{Schmidt2015}, declines to a low level by L4/L5 spectral types and is consistent with this low level through to late T dwarfs. Although the complete sample presented here does not have the virtue of a consistent detection threshold, as the subsample analyzed by \cite{Schmidt2015} does for the L dwarf activity fractions, putting everything together allows for a straightforward comparison between the T dwarfs and the L dwarfs. 

It is clear that the number of objects with H$\alpha$ emission for spectral types later than about mid-L is low. For all of the L dwarfs in this compilation, 67/195, $34\pm^{3.5}_{3.2}$\%, show H$\alpha$ emission. This detection fraction, however, is skewed by the high number of active early L dwarfs. For mid-to-late L dwarfs (L4-L8), only 7/75, $9.3\pm^{4.5}_{2.4}$\%, show H$\alpha$ emission. For comparison, despite nearly doubling the number of measurements available in the literature for T dwarfs, our results show that most T dwarfs show no emission or very weak emission. Only 3/34, $8.8 \pm^{7.4}_{2.8}$\%, distinct systems with T dwarf optical spectral types show H$\alpha$ emission (see Table~\ref{tab:Halpha} and Table~\ref{tab:HalphaL}). Luhman 16B, the nearby T dwarf, also has an EW limit, EW $<1.5$ \AA, but no flux limit, so we did not include it in Table~\ref{tab:HalphaL} \citep{Faherty2014}. Additionally, the 7 WISE T dwarfs with optical spectra from \cite{Kirkpatrick2011}, but no flux measurements, also do not show any indication of H$\alpha$ emission. Inclusion of these targets leads to the statistic that only 3/42, $7.1\pm^{6.2}_{2.2}$\%, of T dwarf systems show this emission feature. Given the broad similarities between the H$\alpha$ detections of T dwarfs and late L dwarfs, we can also group them together to get an overall detection fraction for optical spectral types L4 - T8 of 10/109, $9.2\pm^{3.5}_{2.1}$\%. Inclusion of the additional 8 T dwarfs without flux limits gives, for L4-T9, a detection rate of 10/117, $8.5\pm^{3.3}_{1.9}$\%. Since, we do not treat binaries separately, these figures could even decrease when accounting for each component in multiple systems. 

Interestingly, these detection rates for late L dwarfs and T dwarfs are comparable to the total detection rate, $\sim$7 \%, in surveys looking for brown dwarf radio emission in objects $\ge$L6 \citep{Kao2016, Lynch2016}. If auroral processes are the dominant mechanisms responsible for magnetic emission in late L dwarfs and T dwarfs, these results suggest that geometric beaming of the radio emission is potentially totally absent or may not significantly affect the auroral detection rates.

\begin{deluxetable}{r c c c } % 5 cols
\tablecaption{New H$\alpha$ Measurements 
\label{tab:Halpha} }
\tablehead{ \colhead{Object} & \colhead{SpT Opt} & \colhead{ $f_{\alpha}$ ($10^{-18}$ erg s$^{-1}$ cm$^{-2}$)} & \colhead{$\log (L_{\mathrm{H\alpha}}/L_{\mathrm{bol}})$}}
\startdata

2MASS 0036+1821\tablenotemark{a} & L3.5  &  ... & $-6.1$ \\ [2pt]
2MASS 0700+3157 & L3.5 	& $<15$   & $<-6.4$\\ [2pt]
2MASS 0835$-$0819 & L5 	& $<12$   	& $<-6.5$ \\ [2pt]
2MASS 1507$-$1627 & L5 	& $<17$   & $<-6.5$ \\ [2pt]
2MASS 1750$-$0016	& L5 	& $21.4 \pm 4.8$   & $-6.2 \pm 0.1$ \\ [2pt]
SDSS 1416+1348 & L6 	& $<18$   & $<-6.6$ \\ [2pt]
WISE 1647+5632 & L7 	& $<4.6$   &  $<-5.6$ \\ [2pt]
SDSS 0423$-$0414 & L7.5  	& $16.3 \pm 1.7$   & $-5.9 \pm 0.1$ \\ [2pt]
SDSS 1052+4422 & L7.5 	& $<4.7$   & $<-5.8$ \\ [2pt]
2MASS 1043+2225 & L8  & $4.7 \pm 1.5 $   & $-5.8 \pm 0.2$ \\ [2pt]
SIMP 0136+0933\tablenotemark{b} & T2 	& $<4.9$   & $<-6.6$ \\ [2pt]
WISE 0656+4205 & T2 	& $<3.1$   & $<-6.0$ \\ [2pt]
2MASS 2139+0220& T2 	& $<4.8$   & $<-6.0$ \\ [2pt]
SDSS 0758+3247 & T3 	& $<9.6$   & $<-5.8$ \\ [2pt]
PSO 247+03 & T3 	& $<4.5$   & $<-6.0$ \\ [2pt]
WISE 0819$-$0335 & T4 	& $<6.6$   	& $<-5.7$  \\ [2pt]
2MASS 1750+1759 &  T4 	& $<13$    &  $<-5.0$ \\ [2pt]
SDSS J0000+2554  & T5 	& $<5.7$   & $<-5.7$ \\ [2pt]
2MASS 0559$-$1404 & T5 & $<5.1$   & $<-6.3$ \\ [2pt]
SDSS 0926+5847 & T5 	& $<4.5$ &$<-5.5$   \\ [2pt]
2MASS 2254+3123 & T5 	& $<4.8$   & $<-5.8$ \\ [2pt]
2MASS 0243$-$2453 & T5.5 	& $<3.8$   & $<-5.7$  \\ [2pt]
2MASS 1754+1649 &  T5.5 	& $<5.1$   & $<-5.4$ \\ [2pt]
2MASS 1231+0847 & T6 	& $<8.4$   &  $ < -5.3$\\ [2pt]
WISE 1506+7027 & T6 	& $<5.8$   & $<-6.0$ \\ [2pt]
SDSS 1624+0029 & T6 	& $<4.0$   & $<-5.7$  \\ [2pt]
2MASS 0727+1710 & T8 & $<4.2$   & $<-5.7$ \\ [2pt]
2MASS 0939$-$2448 & T8 	& $<2.8$   & $<-5.8$  \\ [2pt]
2MASS 1114$-$2618 & T8 	& $<6.8$   & $<-5.4$

\enddata
\tablenotetext{a}{The value of $L_{\mathrm{H\alpha}}/L_{\mathrm{bol}}$ for 2MASS 0036+18 was determined using the measured EW and a $\chi$ value $1.415\times10^{-6}$ from \citep{Schmidt2014}, taking the averages for the median $\chi$ of spectral types L3 and L4}
\tablenotetext{b}{The value listed for this object is only from the DEIMOS spectrum taken on 2014 December 22nd}
\end{deluxetable}

\begin{deluxetable}{r c c c } % 5 cols
\tablecaption{Literature T dwarf H$\alpha$ Emission\tablenotemark{a}
\label{tab:HalphaL} }
\tablehead{ \colhead{Object} & \colhead{SpT Opt} & \colhead{ $f_{\alpha}$ ($10^{-18}$ erg s$^{-1}$ cm$^{-2}$)} & \colhead{$\log (L_{\mathrm{H\alpha}}/L_{\mathrm{bol}})$}}
\startdata
SDSS 0837$-$0000\tablenotemark{b} & T0 & $<4.4$ & $<5.3$ \\ [2pt]
SDSS1254$-$0122 & T2 & $7.5 \pm 2.5$& $-5.9 \pm 0.2$ \\ [2pt]
SDSS 1021$-$0304\tablenotemark{bc} & T4 & $<8.3$ & $<-5.3$ \\ [2pt]
2MASS 1209$-$1004\tablenotemark{bc}  &  T4 & $<1.7$ & $<-6.1$ \\ [2pt]
2MASS 0559$-$1404 &   T5& $<6.1$ & $<-6.2$ \\ [2pt]
2MASS 0755+2212 & T6 & $<12$ & $<-5.1$ \\ [2pt]
2MASS 1225$-$2739 & T6 & $<6.7$& $<-5.5$ \\ [2pt]
2MASS 1503+2525 &  T6& $<9.6$ & $<-5.9$ \\ [2pt]
2MASS 1534$-$2952 & T6& $<17$ & $<-5.3$ \\ [2pt]
SDSS 1624+0029 & T6 	& $<4$   & $<-5.7$  \\ [2pt]
2MASS 0937+2931 & T7 & $<3.9$ & $<-6.2$ \\ [2pt]
2MASS 1047+2124& T7& $5.9 \pm 2.7$ & $-5.5 \pm 0.2$ \\ [2pt]
2MASS 1217-0311 & T7& $<7.7$ & $ <-5.4$ \\ [2pt]
2MASS 1237+6526\tablenotemark{d} & T7 & $74.4 \pm 0.8$ &$-4.2\pm0.1$ \\ [2pt]
SDSS 1346$-$0031 & T7 & $<7$ & $ <-5.2$  \\ [2pt]
GL570D & T7 &  $<6.5$ & $<-5.6$ \\ [2pt]
2MASS 0415$-$0935 & T8 & $<7.9$ & $<-5.5$  \\ [2pt]
2MASS 0727+1710 & T8 & $<3.6$& $<-5.9$ 

\enddata
\tablenotetext{a}{Unless otherwise noted, flux measurements are from \cite{Burgasser2000} or B03}
\tablenotetext{b}{Flux values are newly determined from archival spectra}
\tablenotetext{c}{Optical Spectral types are from this paper, updating values presented in \cite{Kirkpatrick2008}}
\tablenotetext{d}{We report the average for this source taken from \cite{Burgasser2002b}}
\end{deluxetable}

\section{Interesting Individual Objects}\label{sec:objects}

\subsection{2MASS 0036+1821}\label{sec:0036}

This target is one of the few L dwarfs to exhibit detectable quiescent radio emission, as well as periodic highly polarized radio pulses \citep{Berger2002,Berger2005, Hallinan2008,McLean2012}. Consequently, there have been numerous studies examining the magnetic activity of this object, looking for X-ray, radio and H$\alpha$ emission \citep{Berger2005, Hallinan2008,Reiners2008}. Previous studies in the optical report limits on the H$\alpha$ emission of EW $<0.5 \; \mathrm{\AA}$ and $<1.0 \; \mathrm{\AA}$ from \cite{Kirkpatrick2000} and \cite{Reiners2008}, respectively. The most stringent previous limit comes from a 4 hr monitoring  observation by \cite{Berger2005} in which they do not detect anything to a limit of $\log( L_{\mathrm{H\alpha}}/L_{\mathrm{bol}} ) \lesssim -6.7$. As the only radio pulsing brown dwarf to not show H$\alpha$ emission, we decided to observe it further due to the potential association of the radio emission to H$\alpha$ emission. Our new observations on 2012 July 19th (UT), clearly show an emission feature at 6563 $\mathrm{\AA}$ with EW = $0.59 \pm 0.08$ and $\log( L_{\mathrm{H\alpha}}/L_{\mathrm{bol}})$ = -6.1 (see Figure~\ref{fig:M0036}). Since our spectrum of 2MASS 0036+1821 was not flux calibrated, we did not measure the flux of H$\alpha$ emission, instead, we used the revised $\chi$ factors of \cite{Schmidt2014} to convert the measured EW to $\log( L_{\mathrm{H\alpha}}/L_{\mathrm{bol}})$. This measurement is in line with some of the previous limits, however, the detection greatly exceeds the limit placed by \cite{Berger2005}. Although many L dwarfs have been shown to exhibit variable H$\alpha$ emission, as evidenced in the compilation by \cite{Schmidt2015}, the emission is generally not as weak as we have detected for 2MASS 0036+1821, nor have most of these targets been monitored over their full rotational periods. Thus, the intermittent variability that we are detecting, at timescales definitively exceeding the rotational period, represents a new phenomena.

To explain their observed radio emission, \cite{Berger2005} considered the possibility that it could be the result of enhanced activity due to a tidal interaction with a close in companion which orbits on a timescale consistent with the 3 hr period in their data. More recent results, however, positively attribute the radio emission to the ECMI and the periodic signature to a combination of the brown dwarf's rotation and the beaming effect of the emission mechanism \citep{Hallinan2008}. Additionally, the new H$\alpha$ emission suggests the presence of long-term variability to the magnetic processes.

In the context of auroral radio emission and its potential connection to H$\alpha$ emission, the intermittent variability of this object can be coherently explained via a potential flux tube interaction between the brown dwarf and a satellite, whose orbit modulates the long-term H$\alpha$ emission. Energetic electrons moving along the field lines are responsible for the radio pulses and generate the H$\alpha$ emission when they precipitate into the atmosphere and deposit their energy at the flux tube footpoint. This scenario is analogous to the interaction between Jupiter and its moon Io (e.g. \citealt{Vasavada1999}). For this scenario to be consistent with the data, the satellite must orbit with a period $\gtrsim8$ hrs, or else the monitoring campaign of \cite{Berger2005} should have seen some indication of H$\alpha$ emission.

The presence of a potential companion is also consistent with the inclination, $i$, of this system. \cite{Crossfield2014b} report a $\varv \sin i$ of $40.0 \pm 2.0$ km s$^{-1}$, which is a weighted average of the consistent measurements from \cite{Jones2005}, \cite{ZapateroOsorio2006} , and \cite{Reiners2008}. Early efforts to understand the magnetic emission from 2MASS~0036+1821 were confounded by the low $\varv \sin i = 15 \pm 5$ km s$^{-1}$ measurement from \cite{Schweitzer2001}. However, \cite{Reiners2008} attributed that outlying value to mismatches between the observed spectra and the atmospheric models used by \cite{Schweitzer2001}. Using a rotational period of 3.08 hrs, $\varv \sin i$ of 40 km s$^{-1}$ and a radii range between 0.9 $R_{\mathrm{Jup}}$ and 1 $R_{\mathrm{Jup}}$, for this field brown dwarf, gives the range of inclinations, $i \sim 80$-$90^{\circ}$ \citep{Hallinan2008}. During the course of the satellite's orbital evolution, the corresponding flux tube footpoint, the location of the H$\alpha$ emission, traces a path around the magnetic axis of the brown dwarf. Since the magnetic axis is not likely to be very misaligned with the rotational axis (for example, Jupiter's magnetic axis differs by only $\sim10^{\circ}$ from its rotational axis \citealt{Badman2015}), and since the brown dwarf has a high inclination, it is very plausible that a hidden satellite could be modulating the H$\alpha$ for this target, the emission being visible during certain orbital phases but hidden on the opposite face of the brown dwarf during others.

Depending on its orbital semi-major axis and orbital inclination, there is a possibility that such a satellite could be transiting the system. For example, an Earth sized satellite around a Jupiter sized brown dwarf would produce a transit depth of $\delta = (R_{\oplus} / R_{\mathrm{Jup}})^{2} = 0.008$. Photometric monitoring from the ground by \cite{Harding2013} detected rotational variability in two 5 hour observations in $I$ band observations of 2MASS~0036+1821 with 1\% photometric precision, however they did not see any transits. \textit{Spitzer} monitoring with 0.1\% photometric precision also detected variability but no transits in their 14 hr observation \citep{Metchev2015}. These observing campaigns could have missed the transit for a longer period satellite, or the object may not be transiting at all. By comparison, Io orbits Jupiter with an 1.77 day period. If an Earth sized planet is placed in a 1.77 day orbit around a 50 $M_{\mathrm{Jup}}$ brown dwarf of radius $R_{\mathrm{Jup}}$, it would orbit at a distance of about 22 $R_{\mathrm{Jup}}$. The plane of the orbit would need to be inclined at an angle, $i_{p}$, such that the $\cos i_{p} < (R_{\star} + R_{p}) / r$, for the planet to transit (for $i_{p} = 0$, the plane of the orbit coincides with the plane of the sky, face on; \citealt{Winn2010}). Using these orbital parameters gives an inclination of $i_{p} > 87.1^{\circ}$; assuming that all orbital inclination are equally likely gives such a satellite a 3\% probability of transiting. If the orbital inclination is consistent with the rotation axis, as it is in many exoplanetery systems, the chances of transiting are much higher \cite{MortonWinn2014}. The current data is suggestive, but more extensive monitoring is required to confirm whether a satellite body is responsible for the long-term modulation of the H$\alpha$ emission.

\subsection{J1750$-$0016}

2MASS 1750$-$0016 is a L5.5 dwarf discovered by \cite{Kendall2007}. Only recently was this target observed at optical wavelengths by \cite{Burgasser2015} and they place an H$\alpha$ EW emission limit of $<0.4 \; \mathrm{\AA}$. On the other hand, we detect excess emission at the location of H$\alpha$ in our DEIMOS spectrum of this target and measure an emission strength of EW~$=0.46 \pm 0.10 \; \mathrm{\AA}$ (see Figure~\ref{fig:Ha_prof} and Table~\ref{tab:Halpha}). Although this emission is rather weak, \cite{Burgasser2015} report detections of a similar level in some of the other targets in their sample. Our new findings suggest that this target could have variable H$\alpha$ emission like 2MASS~0036+1821 or many of the variable targets compiled by \cite{Schmidt2015}.

\subsection{SDSS 0423-0414AB}\label{sec:M0423}

This target was revealed to be a binary system by \cite{Burgasser2005} in HST NICMOS imaging, with a L6 primary and T2 secondary. The target also showed strong H$\alpha$ emission, with EW = $3\; \mathrm{\AA}$ and strong Li \textsc{i} absorption, with EW~=~$11$~\AA~\citep{Kirkpatrick2008}. We used our new DEIMOS observations to once again measure these features, looking for indications of variability. We measured the H$\alpha$ emission, as described in Section~\ref{sec:Ha}, to be EW = $2.95 \pm 0.30$. This value is consistent with the values reported in the literature, suggesting that the emission may be steady across several year time scales. 

We also compared the different measurements of the Li \textsc{i} absorption. We applied the alkali line fitting from Section~\ref{sec:alkali} to both our new DEIMOS spectrum and the previous LRIS spectrum from \cite{Kirkpatrick2008}. Both spectra yielded consistent results, however they were systematically higher than the reported values in the literature. This is likely due to the fact that the Lorentz line profile includes absorption in the wings of the distribution that may not be included by simply subtracting a continuum and adding up the flux in a predefined region around the line center. For consistency with the literature, we report an EW = $11.1 \pm 0.4$, in line with the literature value.

\subsection{SDSS 1052+4422}

This target had been designated as an early T dwarf (T0.5) by \cite{Chiu2006} in their discovery paper, based on the NIR spectrum. However, \cite{Dupuy2015} showed that SDSS~1052+4422 is actually a binary system from adaptive optics imaging. Their detailed study was able to determine dynamical masses of each component based on astrometric monitoring \citep{Dupuy2015}. They also decomposed the composite NIR spectrum from the IRTF/SpeX library and determine spectral types of L6.5 $\pm$ 1.5 and T1.5 $\pm$ 1.0 \citep{Dupuy2015}. Our new integrated light optical spectrum of this target fit between the L7 and L8 optical standards, and we assigned it a spectral type of L7.5. Our observations are thus consistent with the binary decomposition of L6.5 and T1.5, and provide further constraints on the properties of these objects. Binary systems like these, straddling the L/T transition, are important benchmarks for understand the evolution of brown dwarfs. For a given system, a large discrepancy between the NIR integrated light spectral type and the optical integrated light spectral type can be used as an indicator of a potential binary. This highlights the ability of optical spectra, as a counterpart to the NIR spectra, to be a useful diagnostic in verifying binary systems (see also \citealt{Manjavacas2015}).

\begin{figure}[htbp]
   \centering
   \includegraphics[width=0.5\textwidth]{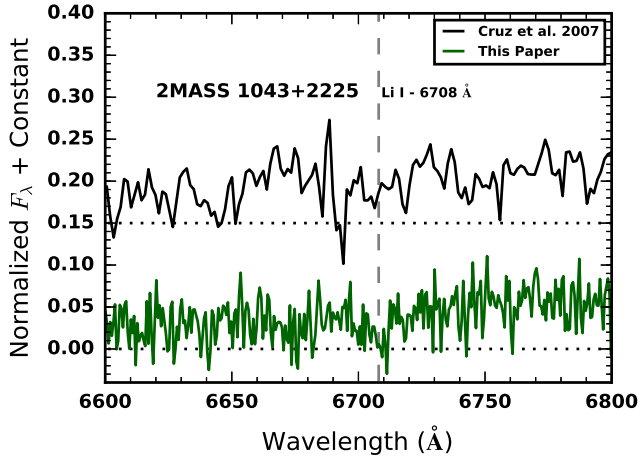} % requires the graphicx package
   \caption{The spectra of 2MASS 1043+2225 zoomed in around the location of the Li \textsc{i} line at 6708 $\mathrm{\AA}$, showing both the spectrum from C07 (top; from the Ultracool RIZzo Spectral Library) and our DEIMOS spectrum (bottom). Our new spectrum shows a dip in the observed flux that could be Li \textsc{i} absorption, however the previous spectrum from C07 only shows  a small trough, in line with the continuum noise. We consider the corresponding lithium detection for this target as tentative.} 
   \label{fig:M1043_Li}
\end{figure}

\subsection{2MASS 1043+2225}

2MASS 1043+2225 is a late L dwarf reported by C07 to have tentative indications of H$\alpha$ emission. Although, they see excess flux at the location of H$\alpha$, their results were inconclusive. Our new observations of this target confirm that this object does indeed exhibit weak H$\alpha$ emission at a level of $\log(L_{\mathrm{H\alpha}}/L_{\mathrm{bol}}) = -5.8 \pm 0.2$ (see Table~\ref{tab:Halpha}). The detection is only just at the 3.1$\sigma$ level, very similar to the weak detections of 2MASS 1047+2124 and 2MASS 1254-0122 from B03. We show the spectrum of this target around the H$\alpha$ line in Figure~\ref{fig:Ha_prof}. 

For this object, we also report a tentative detection of Li \textsc{i} at 6708 $\mathrm{\AA}$. In Figure~\ref{fig:M1043_Li}, we show this region of the spectrum alongside the the spectrum of C07, taken from the Ultracool RIZzo Spectral Library. We measured the absorption to have an EW = $10 \pm 3\; \mathrm{\AA}$, in line with the typical EW of L8 dwarfs with Li \textsc{i} detections \citep{Kirkpatrick2008}. Our more recent, higher resolution, observation shows that there may be an absorption feature there, however, the earlier spectrum does not. We consider this to be a tentative detection which will require deeper observations to confirm. If the absorption is real, this target would be added to the few very late L dwarfs and T dwarfs to display this important physical indicator of mass and age.

\begin{figure}[htbp]
   \centering
   \includegraphics[width=0.5\textwidth]{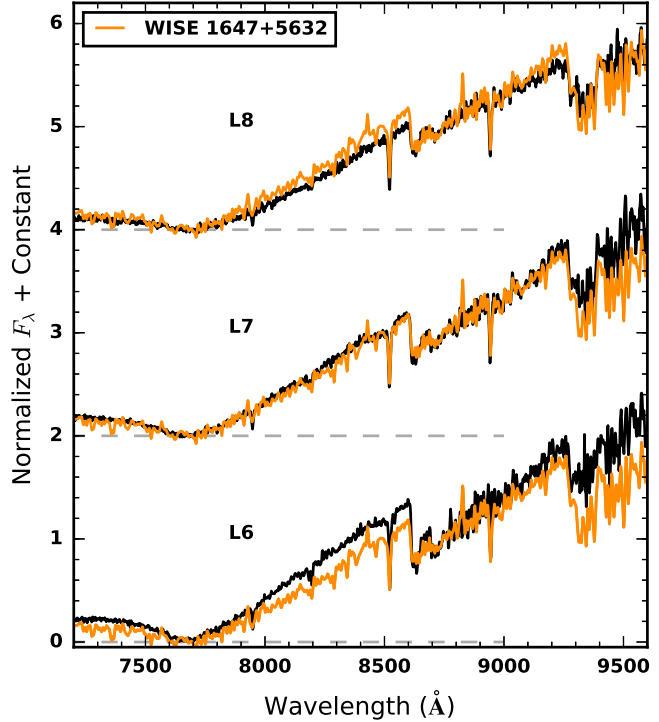} % requires the graphicx package
   \caption{A comparison of the DEIMOS spectrum of WISE1647+5632 (orange) against the optical standards for spectral types L6, L7 and L8 (black). The best match is produced by the L7 standard. This spectral type differs greatly from the NIR spectral type of L9p, and suggests that this target may be an unresolved binary (see Section~\ref{sec:WISE1647})}
   \label{fig:WISE1647}
\end{figure}

\subsection{WISE 1647+5632}\label{sec:WISE1647}

This target is included in the WISE discoveries from \cite{Kirkpatrick2011} and has a preliminary parallax placing it within 10 pc of the Sun. \cite{Kirkpatrick2011} assigned this object a NIR spectral type of L9 peculiar from an IRTF/SpeX spectrum, noting the discrepancies  at H and K band between the spectrum and the standards. They added this object to a collection of unusually red L dwarfs. However, our optical spectrum of this target matches the L7 standard very well (see Figure~\ref{fig:WISE1647}). Our findings suggest that WISE 1647+5632 is likely an unresolved binary system straddling the L/T transition.

\subsection{2MASS 2139+0220}

This target is one of the prominent T dwarfs with high amplitude variability in the J band, displaying up to 26\% variability \citep{Radigan2012}. We included it in the sample to investigate if there could be any connection between the magnetic emissions and the cloud phenomena. We did not find any H$\alpha$ emission and report an emission upper limit of EW $<8 \; \mathrm{\AA}$. There have also been some suggestions that this object could be a binary due to its somewhat peculiar NIR spectrum \citep{Bardalez2014}. Our observed spectrum matches the T2 optical spectral standard very well. This corroborates findings by \cite{Manjavacas2015} which rule out this scenario. The peculiar spectrum is thus, likely a consequence of a patchy atmosphere and the impact of cloud variability on the emergent spectral flux. 

\subsection{SIMP 0136+0933}\label{sec:SIMP}

SIMP 0136+0933 is one of the archetypes for cloud variability at the L/T transition; it was found to exhibit 50 mmag photometric variability in J band and has since been followed up throughout the infrared to characterize the patchy cloud structures of its atmosphere \citep{Artigau2009, Apai2013, Radigan2014}. As a potentially very interesting object in the context of auroral activity, we observed it with LRIS on 2014 August 27th, and again with DEIMOS on 2014 December 22nd. In the first epoch we took two consecutive 1200 s exposures, whereas in the second epoch we took two exposures of 1800 s each, separated by 1.5 hrs. 

In no exposure did we detect any excess flux at the location of the H$\alpha$ line. We measured stringent limits on the corresponding emission line flux of SIMP~0136+0933 from the co-added DEIMOS spectrum (see Table~\ref{tab:Halpha}). Because we detect the underlying continuum in the combined spectra, for this target, we also report EW emission limits of $<3.2\; \mathrm{\AA}$ and $<3.5\; \mathrm{\AA}$ for the August and December nights, respectively. In the context of auroral emission, which may be rotationally modulated, the 1.5 hrs of separation between the exposures in December correspond to a phase shift of 0.63, using the photometric rotational period of 2.39 hrs \citep{Apai2013}. Although, it remains possible that we missed potential optical auroral emission from this source, the series of observations at different phases suggests that it may indeed lack H$\alpha$ emission. 

\footnotetext[7]{Disentangling the effective temperature, gravity and cloud effects remains a challenging problem in brown dwarf atmospheric modeling and depending on the different assumptions can yield differing answers, even when good data are available \citep{Marley2010, Apai2013, Faherty2014, Marley2015}. }

\begin{figure}[htbp]
   \centering
   \includegraphics[width=0.5\textwidth]{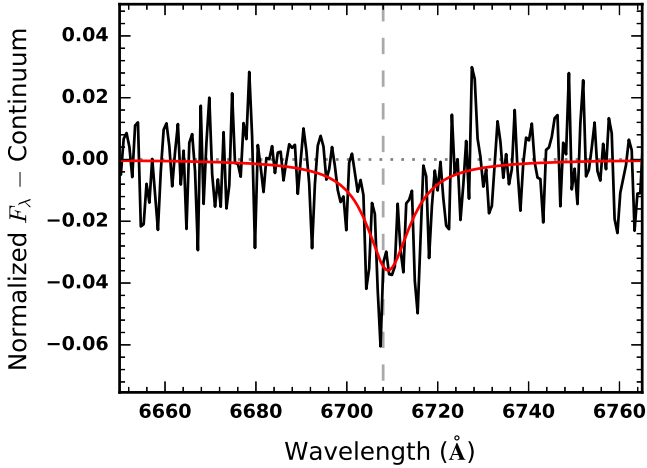} % requires the graphicx package
   \caption{The spectrum of SIMP 0136+0933 around the Li \textsc{i} line at 6708 $\mathrm{\AA}$. The spectrum is normalized with the continuum profile subtracted. The line corresponds to a Lorentz line profile model fit of the data (see Section~\ref{sec:alkali}). }
   \label{fig:exSIMP_Li}
\end{figure}

Our high signal-to-noise ratio spectrum, S/N $\sim6$ at 6800~\AA~and S/N $\sim96$ at 8600~\AA, of SIMP~0136+0933 also allowed us to look for the presence of Li at 6708 $\mathrm{\AA}$. We plot this spectrum in Figure~\ref{fig:exSIMP0136} with the inset zoomed in on the location of the Li \textsc{i} absorption. It is clearly present. We fit the absorption line as we fit the other alkali lines in Section~\ref{sec:alkali} with a Lorentz line profile and over-plotted the model result in Figure~\ref{fig:exSIMP_Li}. As in the case of SDSS 0423-0414, we report EW values not based on the fit but a simple summation of the absorption line region (see Section~\ref{sec:M0423}). We measure EW values of $6.6 \pm 1.0\; \mathrm{\AA}$ and $7.8 \pm 1.0\; \mathrm{\AA}$ for the August and December observations respectively. SIMP 0136+0933 joins the T0.5 dwarf, Luhman 16B (EW $= 3.8 \pm 0.4$ ), as the second T dwarf and the latest spectral type object with a clear lithium detection in its atmosphere \citep{Faherty2014}. Although the spread of the values between the two objects is in line with the spread of detections for L8 dwarfs from \cite{Kirkpatrick2008}, this absorption appears to be particularly strong by comparison given that SIMP 0136+0933 has a later spectral type and possibly cooler atmosphere\footnotemark[7]. As \cite{Lodders1999} discussed, the Li \textsc{i} in the atmosphere becomes readily depleted by the formation of LiCl gas and other Li bearing substances like LiOH in cool dwarf atmosphere below temperatures of about 1600 K. \cite{Kirkpatrick2008} showed that the peak of Li \textsc{i} absorption takes place around a spectral type of L6.5 and declines for later spectral types. 

\begin{figure}[htbp]
   \centering
   \includegraphics[width=0.5\textwidth]{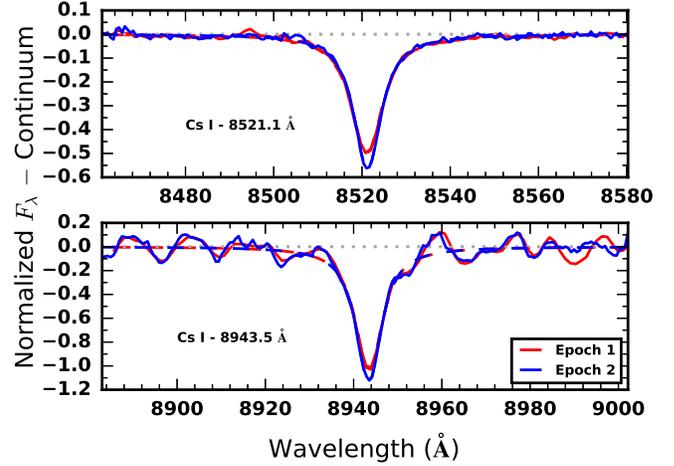} % requires the graphicx package
   \caption{The Cs \textsc{i} absorption lines of SIMP~0136+0933 between the observations in August and December plotted as solid lines and the Lorentz line profile model fits plotted as dashed lines. The pEWs of the lines are significantly different and likely reflect spectroscopy variability between the two observations for this photometrically variable target, see Section~\ref{sec:SIMP}}
   \label{fig:CsComp_SIMP}
\end{figure}

\footnotetext[8]{We did not see any indication of Li \textsc{i} in our LRIS spectrum of 2MASS 2139+0220, however there was not enough signal in the continuum to definitively rule it out.}

Thus, the strong Li \textsc{i} of SIMP 0136+0933 is somewhat anomalous, however it is interesting to note that in addition to SIMP 0136+0933, Luhman 16B also exhibits cloud variability \citep{Crossfield2014,Burgasser2014}. The presence of Li may be related to the transition from L to T spectral types within a patchy cloud atmosphere \footnotemark[8]. Indeed, although our Li \textsc{i} EW measurements from the different epochs are formally in agreement within the 2$\sigma$ level, the central values differ by 15\%. If this is due to cloud phenomena in the atmosphere, the spectra from the  different epochs could be dominated by flux from atmospheric levels with differing Li \textsc{i} depletion. 

The two different epochs did allow us to observe spectroscopic variability in the other optical absorption lines. We note that the difference of the pEWs in the co-added spectra from the two epochs, for the lines reported in Table~\ref{tab:alkali}, are statistically significant. This is especially true of the Cs \textsc{i} lines where the signal-to-noise ratio is greatest. For the first epoch, 2014 August 27th, we measured pEWs of $10.16\pm0.06$~\AA~and $8.45\pm0.04$ for the Cs \textsc{i} lines at 8521~\AA~and 8943~\AA~respectively. For the second epoch, 2014 December 22nd, we measured pEWs of $9.70\pm0.06$~\AA~and $8.58\pm0.04$ for the same lines, respectively. The difference between the pEW measurements for the Cs \textsc{i} line at 8521 \AA~is different from 0 at the 5.4$\sigma$ level and similarly at the 2.3$\sigma$ level for the Cs \textsc{i} line at 8943 \AA. The Rb \textsc{i} lines at 7948 \AA~were measured to have pEWs of $10.96\pm0.38$ \AA~and $9.96\pm0.32$ \AA~for epoch 1 and epoch 2, respectively, yielding a difference that is significant at the 2$\sigma$ level. 
In Figure~\ref{fig:CsComp_SIMP}, we plot a comparison of these spectral line profiles with the continuum subtracted and the corresponding Lorentz profile model fits (see section~\ref{sec:alkali}).

The pEW measurements track the changes in the absorption relative to the nearby pseudo-continuum. The two different Cs \textsc{i} line observations did not show the same degree of variation, suggesting that this variability may be driven as much by differences in the relative continuum in the different parts of the spectrum as in the individual absorption line strength. These results provide support for the interpretation of cloud variability in the atmosphere of this object and supports the idea that there could be significant optical variability to coincide with the large-amplitude NIR variability, potentially even in the Li \textsc{i} absorption. In fact, \cite{Heinze2015} showed that photometric optical variability in brown dwarfs could be stronger than the NIR variability and \cite{Buenzli2015} used \textit{Hubble Space Telescope} grism observations to demonstrate spectroscopic variability from 0.8 $\mu$m to 1.15 $\mu$m in both components of Luhman16AB.

\section{Discussion and Summary}\label{sec:summary}

We have conducted a new survey at red optical wavelengths (6300 $\mathrm{\AA}$ - 9700 \AA) looking for H$\alpha$ emission in a large sample of late L dwarfs and T dwarfs. We acquired new optical spectra for 18 targets without previous spectra and several additional spectra looking for potential variability in the emission features. We have nearly doubled the number of red optical spectra available for T dwarfs and used our new observations, in conjunction with available spectra, to examine prominent spectral features and the optical T dwarf sequence. 

Our findings include two objects that fill the gap between the T dwarf optical spectral standards from T2 - T5. We proposed PSO 247+03 as the T3 spectral standard and WISE~0819-0335 as the T4 spectral standard. These two targets are relatively bright and are both near the equatorial plane, allowing for observational access from both the northern and southern hemispheres.

We also observed Li \textsc{i} absorption at 6708 $\mathrm{\AA}$ in the spectrum of SIMP 0136+0933, one of the most prominent IR photometric variable brown dwarfs. This object becomes only the second T dwarf and the latest type object to display this feature. We also see spectroscopic variability in the strength of the absorption lines that is likely related to the heterogeneous cloud phenomena present in the atmosphere.

Our survey included new H$\alpha$ detections for 2MASS~0036+1821, 2MASS~1750$-$0016, and 2MASS~1043+2225 and many more limits on the H$\alpha$ flux for late L dwarfs and T dwarfs (see Section~\ref{sec:Ha}). Our focus on these objects has allowed us to investigate the prevalence of magnetic activity in objects with low temperature atmospheres. The persistent magnetic emissions of many objects in this regime and the discovery of continued activity, even in late T dwarfs points to deficiencies in the understanding of magnetic atmospheric processes and/or new phenomena that fall outside of the standard paradigm of stellar activity. 

For the warmer UCDs, chromospheric emission may still persist. Recent work by \cite{RodriguezBarrera2015} on the ability of UCD atmospheres to become magnetized suggests that the plasma conditions may allow for objects to remain magnetized down to an $T_{\mathrm{eff}} \sim 1400$ K, 900 K cooler than the similar magnetization threshold considered by \cite{Mohanty2002}. This lower threshold is similar to the typical effective temperatures of L4/L5 dwarfs \citep{Kirkpatrick2005} and would coincide with where we see the detection of H$\alpha$ emission bottom out (see Figure~\ref{fig:DetectFrac}). However, for even cooler objects, the strong optical and radio emissions of some objects remain difficult to explain.

The emergence of the ECMI as a coherent explanation for the periodic radio emissions of numerous studies across the UCD regime provides an alternative. These studies argue that auroral processes are capable of driving the periodic radio and optical emissions that have been observed and are also consistent with potential long-term variability (see Section~\ref{sec:0036}). The benchmark objects that have been used to investigate these processes have predominantly been systems of either late M or early L spectral types. These kinds of objects might exhibit both auroral and/or chromospheric emissions, requiring detailed study to distinguish. This confusion is alleviated when examining the population of late L dwarfs and T dwarfs with atmospheres for which standard Solar-like magnetic processes have difficulty generating chromospheric emissions, due to the highly neutral atmospheres. If the H$\alpha$ emission in these objects is connected to the radio auroral emission, then the prevalence of this emission provides an estimate of the overall occurrence rate of auroral activity.

Our measurements of T dwarf H$\alpha$ emission revealed that this activity indicator is less common than previously thought. B03 found three T dwarfs in about a dozen objects to exhibit this emission, two weak emitters and 1 very strong emitter. Our new observations and other work since their initial efforts show that the emission in this regime is actually much rarer and likely only seen in $\sim$7\% of T dwarf systems. When considering objects of spectral type from L4-T8, the detection rate remains only $9.2\pm^{3.5}_{2.1}$~\% (as low as $8.5\pm^{3.3}_{1.9}$\% for L4-T9). It is possible that some of these targets exhibit variability and we did not observe the targets at the right point in time to catch the emission, however that is unlikely to be the case for all of the targets. Nevertheless, only extended monitoring of each target will be able to rule out that scenario. 

Even if the occurrence rate of auroral activity is well characterized by our H$\alpha$ detection rate of $\sim$10 \%, the question of the nature of the underlying mechanism that governs the emission still remains. \cite{RodriguezBarrera2015} point out that, despite having less magnetized atmospheres, objects with $T_{\mathrm{eff}}<1400$ K are capable of sustaining significant ionospheres and driving auroral emission processes. \cite{Schrijver2009} and \cite{Nichols2012} point to a rotation dominated magnetospheric-ionospheric coupling current system as the underlying mechanism for the auroral emissions capable of generating strong surface emission features near the magnetic poles. However, what determines whether an object displays auroral emission or not? One clue might be the long-term variability we have detected in the H$\alpha$ emission of 2MASS~0036+1821. Within the auroral context, if this emission is proved to be related to the presence of satellites, then our observed detection rate for late L dwarfs and T dwarfs may reflect the satellite occurrence rate.

Comparing our overall H$\alpha$ detection rate to surveys of brown dwarf radio emission revealed that radio and H$\alpha$ detection rates in late L dwarfs and T dwarfs are comparable, suggesting that if the emission is auroral then geometric beaming may not play a prominent role in the detectability of the radio aurorae. Consequently, the sample of H$\alpha$ emitting brown dwarfs are potentially excellent targets to pursue with sensitive radio telescopes, like the \textit{Jansky Very Large Array}. These magnetically active brown dwarfs will be important benchmark objects for understanding not only magnetospheric processes across the brown dwarf regime from planets to stars but also for understanding magnetic dynamos in fully convective objects \citep{Kao2016} and examining the upper atmospheric structures of brown dwarfs.

\section*{Acknowledgments}

The authors wish to recognize and acknowledge the very significant cultural role and reverence that the summit of Mauna Kea has always had within the indigenous Hawaiian community.  We are most fortunate to have the opportunity to conduct observations from this mountain.

The authors would like to thank the anonymous referee for helpful comments strengthening this contribution. J.S.P would like to thank Yi Cao for assistance in DEIMOS data reduction. J.S.P was supported by a grant from the National Science Foundation Graduate Research Fellowship under grant No. (DGE-11444469). 

This research has benefitted from the M, L, T, and Y dwarf compendium housed at DwarfArchives.org. This research has benefitted from the Ultracool RIZzo Spectral Library maintained by Jonathan Gagn\'{e} and Kelle Cruz. This researched has benefitted from the Database of Ultracool Parallaxes maintained by Trent Dupuy.

PyRAF is a product of the Space Telescope Science Institute, which is operated by AURA for NASA. PyFITS is a product of the Space Telescope Science Institute, which is operated by AURA for NASA.

The data presented herein were obtained at the W.M. Keck Observatory, which is operated as a scientific partnership among the California Institute of Technology, the University of California and the National Aeronautics and Space Administration. The Observatory was made possible by the generous financial support of the W.M. Keck Foundation.

This publication makes use of data products from the Two Micron All Sky Survey, which is a joint project of the University of Massachusetts and the Infrared Processing and Analysis Center/California Institute of Technology, funded by the National Aeronautics and Space Administration and the National Science Foundation.

\bibliographystyle{apj}
\bibliography{ms}

\end{document}